\documentstyle[aps,feynmf,epsfig]{revtex}

\begin{document}
\draft

\title{Electronic topological  transition in 2D electron system on
a square lattice
 and the line $T^*(\delta)$ in the underdoped regime of high-$T_c$ cuprates}

\author{F. Onufrieva \and P.Pfeuty}
\address{Laboratoire Leon Brillouin CE-Saclay 91191 Gif-sur-Yvette France}

\normalsize

\maketitle

\date{\today}

\begin{abstract}

It is shown
that a 2D system of free fermions on 
a square lattice with hoping between more than nearest neighbours
undergoes   a fundamental
 electronic topological
transition (ETT) at some
electron concentration $n_c$.
 The point $\delta=\delta_c$, $T=0$ ($\delta=1-n$)
 is an exotic  quantum critical
point with  several  aspects of criticality.
 The  first trivial one is related to
 singularities in thermodynamic
properties.
An  untrivial and never considered aspect is
related to the effect of Kohn singularity (KS) in 2D system 
 : this point   is  the end
of a critical line $T=0$, $\delta>\delta_c$
of   static
KS in free fermion polarizability.
 This ETT is a motor for anomalous behaviour in the
system of interacting electrons. The anomalies take place
on one side of ETT,   $\delta<\delta_c$, and
 have a striking similarity
with the anomalies in the 
high $T_c$ cuprates in the underdoped regime. 
The most important consequence of ETT is the appearance of
the line 
of characteristic temperatures, $T^*(\delta) \propto 
\delta_c-\delta$, which grows from the point $\delta=\delta_c$
on the side
$\delta<\delta_c$. Below this line the metal state is anomalous.
Some anomalies are considered in the present paper.
The  anomalies disappear in  the case
  $t'=t''=... \rightarrow 0$.
\end{abstract}

\begin{fmffile}{flo1}

\fmfcmd{%
 style_def anomal expr p =
 cdraw p;
 cfill (tarrow (p, .55));
 cfill (harrow (p, .55));
 enddef;
 style_def wiggly_arrow expr p =
  cdraw (wiggly p);
  shrink (2);
  cfill (arrow p);
  endshrink;
  enddef;}
\section{Introduction}

Many experiments performed for high $T_c$ cuprates
provide an evidence for unusual behaviour in the
underdoped regime. Analysis of NMR \cite{NMR,Takigawa}, angular
photoemission study \cite{ARPES}, infrared conductivity 
\cite{conductivity},
transport properties \cite{transport},
thermopower \cite{thermopower}, heat capacity
\cite{heatcapacity}, 
 spin susceptibility \cite{susc}
and Raman spectroscopy \cite{Raman}
have revealed  the existence of a
characteristic energy scale $T^*(\delta)$ in
the normal state with an absolute value  different
for different properties while universal 
feature is increase of $T^*(\delta)$ with
decreasing doping.

We don't know any theory which predicts the
existence of such a line apart from theories containing
only different hypothesis not explicit calculations.

With this paper we start a series of articles in which firstly
we show the existence of such a line in 2D system of
electrons  on a
square lattice with hoping between more than nearest
neighbours and analyse its origin and then we
demonstrate its appearance for different
properties. In the present paper we consider only the
origin of the line and its appearance in
magnetic properties : SDW correlation length,
NMR and neutron scattering. Electronic properties
are considered in \cite{Kisselev,OPDis}. Transport properties are
currently under study.

We show that when varying the 
electron concentration defined as $1-\delta$,
the system of noninteracting electrons 
on a square lattice undergoes an electronic topological
transition (ETT) at a critical value $\delta=\delta_c$. 
The corresponding quantum critical point
(QCP) has a triple nature. 
The first aspect
of the criticality is related to the local change in FS at 
$\delta=\delta_c$. It leads to  
 divergences in the thermodynamic
properties,
in the density of states at $\omega=0$
(the Van Hove logarithmic singularity \cite{Van-Hove}), in the ferromagnetic (FM)
 response function and to the additional divergence in the
  superconducting
(SC) response function of noninteracting
electrons. This induces  FM and SC instabilities in the
vicinity of QCP in the presence of the interactions of necessary
signs.

The second aspect is related to the topological change in 
 mutual properties of the FS in the vicinities of two 
different SP's and leads to the logarithmic divergence
of spin density and charge density susceptibilities  of noninteracting
electrons. This divergence has an "excitonic" nature
in the sense that the topology of two electron 
bands around two different SP's is such that no energy is needed
to excite an electron-hole pair. This can lead to 
spin-density wave (SDW)  or charge-density wave (CDW)
instability in the presence of the corresponding interaction.

 The change in mutual topology of FS around SP's is
also responsible for the
third aspect of the criticality  related
 to the fact that this point is the end
of a critical line of   static
Kohn singularities in the density-density susceptibility
 which exists only on one
side of QCP, $\delta>\delta_c$.
[What we mean as the static Kohn singularity is a
square-root singularity at $\omega=0$ and wavevector
${\bf q}_m$ which connects two points of Fermi surface (FS) with
parallel tangents \cite{Kohn,Kohn2,Rice}. For an isotropic FS and for 1D case this
wavevector is nothing but "$2k_F$".]

 The latter aspect of criticality exists only in the case
of finite $t'$ or/and  $t''$ etc.
(where $t'$,  $t''$ are hoping
parameters for   next nearest, next next nearest neighbors etc.)
and  has never been considered \cite{neighbours}
However as we show in the
present paper it is just
this aspect that leads to numerous anomalies in the 2D electron
system on the other side of QCP,  $\delta<\delta_c$ which
 have a striking similarity
with the anomalies in the 
high $T_c$ cuprates in the underdoped regime and are in our
opinion at the origin of the latter anomalies.

The system of nonineracting electrons  is considered
in Sec.II. We demonstrate the existence of ETT for
the case of electrons on a square lattice
with $t$ and $t'$ hoping parameters and discuss its
different aspects.
 ETT firstly has been considered  by 
I. Lifshitz \cite{Lifshitz}  for 3D
metals. It was shown that  ETT implies an anomalous behavior and
 singularities
in the thermodynamic and  kinetic properties close to the
transition point \cite{ETT}. We show that for the 2D system the
anomalies are still there but the situation is different
and much more rich. We study the behaviour of density-density
susceptibility and show that it behaves qualitatively
different on two sides of ETT : quite ordinary on one
side, $\delta>\delta_c$, and anomalously on the other,
$\delta<\delta_c$. 
We discover the
existence of a  line $T^*(\delta) \propto \delta_c-\delta$
of Kohn anomalies  which grows from the QCP on the side
$\delta<\delta_c$. These anomalies are related to
dynamic Kohn singularities at $T=0$.

The anomalies in the system of noninteracting electrons
are at the origin of anomalies which appear in the presence
of interaction, see
 Sec.III. 
 Dependently on the type of interaction,
 CDW or SDW instability takes place around QCP. We consider
 a spin-dependent
interaction of the AF sign and respectively the SDW phase.
 Strange about this phase is that the 
critical line  has an unusual shape as a function of $\delta$
in the regime $\delta<\delta_c$ :
{\bf It increases
with increasing the distance from QCP}
instead of having the form  of
a "bell" around QCP as it is usually happens
for an ordered phase developing around an ordinary quantum critical
point and as indeed it occurs on the side $\delta>\delta_c$.
The reason is that {\bf the ordered
phase develops rather "around" the line $T^*(\delta)$
than around the quantum critical point at $T=0,Z=0$}.
 We  call this instability the {\bf SDW
"excitonic" instability}  in order to distinguish it
from the SDW instability associated with nesting
of Fermi surface occuring in the case $t'=t''=...0$. In the
latter case all anomalies discovered in the paper 
disappear. The reason is discussed in \cite{OPQCP2}.

The line  $T^*(\delta)$ 
plays a very important role also for the disordered metallic
state out of the SDW
"excitonic" ordered phase. We show that it is a line
{\bf of almost phase transition}. The correlation length increases
when approaching this line at fixed doping from high and {\bf from
low} temperature, relaxation processes slow down etc.

The metallic state below the line  $T^*(\delta)$
is  quite strange. On one hand,
it exhibits a reentrant behaviour becoming more
rigid with increasing
temperature. On the other hand,
dependences of the correlation length and of the
parameter $\kappa^2_{exc}$ (which describes a
proximity to the ordered SDW "excitonic"
phase) on doping are extremely weak. It
leads to the unusual behaviour : in this regime the system
remains effectively in a  proximity of the ordered SDW
 phase  for all dopind
levels even quite far from $\delta_{exc}(T=0)$.
Therefore the metallic state below the line  $T^*(\delta)$
and out of the  ordered SDW
"excitonic"  phase is a {\bf reentrant in temperature and
almost frozen in doping rigid SDW liquid}. This state
is characterized by numerous anomalies, some of them
are considered in the paper. The existence of this
state frozen in doping
 has very important consequences
when considering the situation
in the presence of SC phase. Since the line
 $T=T^*(\delta)$ is always leaning out of the SC phase,
 the normal
state at $T_{sc}(\delta)<T<T^*(\delta)$ is quite strange :
 Being located in a proximity of the
 SC phase in fact
it keeps a memory about the 
SDW "excitonic" phase. 
It is this feature which on our opinion is crucial for
understanding the anomalous behaviour in the underdoped
regime of high-$T_c$ cuprates.

In the end of the Sec.III we discuss briefly an anomalous
behaviour of physical characteristics corresponding to
those measured by NMR and inelastic neutron scattering (INS). The
 discovered
anomalies are in a good agreement with nuclear spin
 lattice relaxation rate $1/T_1T$  and nuclear transverse
relaxation rate
$1/T_{2G}$ experimental data on copper (see for example
\cite{Takigawa}). We discuss also a
possible reason for the different behaviour of $1/T_1T$
on copper and oxygen \cite{Walstedt}. We do some predictions for INS
which can be verified by performing an energy scan in
the normal state in a progressive change of temperature.
More detailed analysis of the behaviour of 
$Im\chi({\bf q},\omega)$ measured by INS is performed in \cite{OPINS}.
The Sec.IV contains summary and discussion.

\section{Quantum critical point and electronic
 topological transition
in a system of noninteracting 2D electrons}

Let us consider noninteracting  electrons on a quadratic lattice
with nn and nnn hopping. The dispersion law is written as

\begin{equation} 
\epsilon_{{\bf k}\sigma} =
\ -2t (cos k_x + cos k_y ) - 4t' cos k_x cos k_y\
\label{1}
\end{equation}
where $t$ and $t'$ are hoping parameters for nearest and next
nearest neighbors. We consider $t>0$ and $t'<0$ in order to correspond to the experimental situation which reveals the open
FS in the underdoped regime.
  The dispersion (\ref{1}) is characterized 
 by 2 different saddle points (SP's) located at
 $ ( \pm\ \pi, 0)$ and $(0, \pm  \pi)$
  (in the first Brillouin zone $ (-\pi,0)$ is equivalent to $ (\pi,0)$ and
$ (0,-\pi)$ is equivalent to $ (0,\pi )$)  with the energy 
 \begin{equation}
\epsilon_{s} = 4 t' .
\label{2}
\end{equation}
When we vary the chemical potential $\mu$ or the energy
 distance from
the SP, $Z$, determined as 

\begin{equation}
Z=\mu-\epsilon_{s}-=\epsilon_{F}-4 t',
\label{3}
\end{equation}
 the topology
of the Fermi surface changes when $Z$ goes from $Z<0$ to $Z>0$ through
the critical value $Z=0$. In the first Brillouin zone, the Fermi surface
is closed for $Z<0$ and open for $Z>0$. When $Z$ initially negative goes 
positive there is formation of two necks at ${\bf k}=(0,\pi)$ and
 ${\bf k}=(\pi,0)$.
When the Brillouin zone is extended, the Fermi surface goes from 
convex ($Z<0$) to concave ($Z>0$), see Fig.1.

\begin{figure}
\begin{center}
\epsfig{%
file=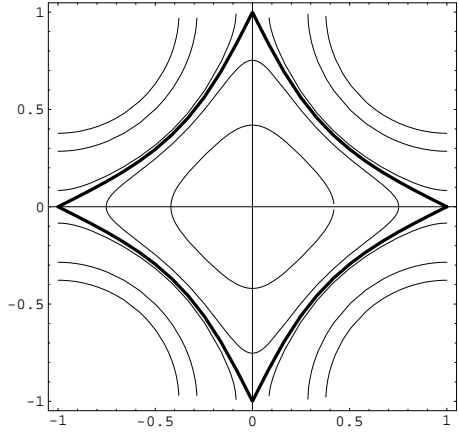,%
figure=figure1.eps,%
height=6cm,%
width=6cm,%
angle=0,%
}
\\
\end{center}
\caption{Fermi surface of the electron system with the dispersion
law (\ref{1}) for different Z and $t'/t=-0.3$. The thick line corresponds to $Z=0$. Here $\tilde{k}_x=k_x/\pi$,   $\tilde{k}_y=k_y/\pi$}
\label{f1}
\end{figure}

 The point $T=0,Z=0$ corresponds to
an ELECTRONIC TOPOLOGICAL TRANSITION (ETT).
It is a quantum critical point
(QCP) which has a triple nature.

\subsection{Three aspects of criticality of the point
$Z=0$, $T=0$}

1. The first aspect  of criticality is related to
{\it  properties of one
SP, i.e. to the local change in topology of FS in
the vicinity of SP}. It results in
singularities in thermodynamical properties, in
FM response function, in  density of states
at $\omega=0$,
 in additional (with respect
 to usual metal) singularity in SC
response function.

The thermodynamical
and FM singularities are given by

\begin{equation}
 \delta G \propto Z^{\frac{d+2}{2}}, \hskip 0.6 cm
\frac{C}{T} =  \frac{d^{2}\delta G}{d T^{2}}
\propto \ ln Z, \hskip 0.6 cm
\chi= \frac{d^{2}\delta G}{d h^{2}} \propto \ ln Z
\label{4}
\end{equation}
as $T \rightarrow 0$ and by

\begin{equation}
 \delta G \propto T^{\frac{d+2}{2}}, \hskip 0.6 cm
\frac{C}{T} =  \frac{d^{2}\delta G}{d T^{2}}
\propto \ ln T, \hskip 0.6 cm
\chi= \frac{d^{2}\delta G}{d h^{2}} \propto \ ln T
\label{5}
\end{equation}
as $Z \rightarrow 0$ (d is dimension, $G$ is the Gibbs
potential, $C$ is heat capacity, $\chi$ is uniform
susceptibility; the expressions
for $\frac{C}{T}$ and $\chi$ are given for $d=2$)
[More generally the singularities in the
thermodynamic properties are described by the expression
 $\delta G \propto Z^{(d+z)\nu} F(\frac{T}{Z^{\nu z}})$
with $z=2$, $\nu=1/2$
( a gaussian type behaviour). We will discuss this elsewhere.]
 
The SC response function for noninteracting electrons is
related to the  Green function

\begin{equation}
G_{sc}^0({\bf k}, i\omega_n)=-T\sum_{{\bf q},\omega_m}
K_{i}({\bf q}, i\omega_m) K_{i}({\bf k-q}, i\omega_{n-m})
\label{6}
\end{equation}
and is given by

\begin{equation}
{\sf \Pi}^0({\bf k},\omega)=-\lim_{\delta \rightarrow 0} G^0_{sc}({\bf k},\omega+i\delta)=
\frac{1}{N}\sum_{{\bf q}}\frac{1-n^{F}(\tilde{\epsilon}_{{\bf q}})
-n^{F}(\tilde{\epsilon}_{{\bf {q+k}}})}
{\tilde\epsilon_{\bf q}
+
\tilde{\epsilon}_{\bf q+k}
- \omega- i0^+}.
\label{7}
\end{equation}
It diverges double logarithmically as a function of $T$
as ${\bf k} \rightarrow 0$,
$\omega \rightarrow 0$ and  $Z=0$ :

\begin{equation}
{\sf\Pi}^{0}({\bf k}=0,\omega=0) \propto
\ln^2T
\label{8}
\end{equation}
When  $T \rightarrow 0$ and $Z  \rightarrow 0$
 it diverges as
\begin{equation}
{\sf\Pi}^{0}({\bf k}=0,\omega=0) \propto  N(Z)
\ln{\frac{\omega_{max}}{T}} \propto
\ln{\frac{\omega_{max}}{T}}
\ln{\frac{\omega_{max}}{|Z|}}.
\label{9}
\end{equation}
where $ N(Z)$ is the density of states.\\

2. The  second aspect is related to mutual
 properties of two different SP's and reveals
itself when considering the density-density
susceptibility (or by other words the electron-hole
response function) at ${\bf k=Q}_{AF}=(\pi,\pi)$,
i.e. at the wavevector which
joins two SP's. This response function
 is related to the electron-hole Green function of
noninteracting electrons 

\begin{equation}
G_{e-h}^0({\bf p}, i\omega_n)=-T\sum_{{\bf q},\omega_m}
K_{1}({\bf q}, i\omega_m) K_{2}({\bf p+q}, i\omega_{n+m})
\label{10}
\end{equation}
and is given by

\begin{equation}
\chi^0({\bf k},\omega)=-\lim_{\delta \rightarrow 0} G^0_{e-h}({\bf k+Q}_{AF},\omega+i\delta)=
\frac{1}{N}\sum_{{\bf q}}\frac{n^{F}(\tilde{\epsilon}_{{\bf q}})
-n^{F}(\tilde{\epsilon}_{{\bf {q+k}}})}{\tilde{\epsilon}_{{\bf q+k}
}-
\tilde{\epsilon}_{{\bf {q}}}-\omega-i0^+}.
\label{11}
\end{equation}
It diverges
logarithmically
 as ${\bf k} \rightarrow {\bf Q}_{AF}$,
 $\omega \rightarrow 0$ and   $Z =0$ :

\begin{equation}
\chi^{0}({\bf Q}_{AF},0) \propto \ln\frac{\omega_{max}}{T},
\label{12}
\end{equation}
and as ${\bf k} \rightarrow {\bf Q}_{AF}$,
 $\omega \rightarrow 0$, $T = 0$ :

\begin{equation}
\chi^{0}({\bf Q}_{AF},0) \propto \ln\frac{\omega_{max}}{|Z|}.
\label{13}
\end{equation}
The one-electron Green functions $K_{i}({\bf k},
\omega_n)$
in (\ref{6}), (\ref{10}) are defined as

\begin{equation}
K_{i}({\bf k},
\omega_n)=\frac{1}{i\omega_n-\tilde{\epsilon}_i({\bf k})}
\label{14}
\end{equation}
where $\tilde{\epsilon}_1({\bf k})=\epsilon_1({\bf k})-\mu$ and
$\tilde{\epsilon}_2({\bf k})=\epsilon_2({\bf k})-\mu$
are the electron spectra
in the vicinity of two different SP's : 

$$\tilde{\epsilon}_1(\bf k) = \tilde{\epsilon}(\bf k),$$
\begin{equation}
\tilde{\epsilon}_2({\bf k}) = \tilde{\epsilon}({\bf k+Q}_{AF}),
\label{15}
\end{equation}
and $\epsilon(\bf k)$ is determined by (\ref{1}) 
(here and below we omitt the index $\sigma$ in the electron
spectrum due to its degeneracy). In the hyperbolic
approximation one has

$$
\tilde{\epsilon}_1({\bf k})/t = -Z/t+a k_x^2 -
b k_y^2, \hskip 1 cm (a=1-2t'/t, \hskip 0.3 cm b=1+2t'/t)
$$
\begin{equation}
\hskip - 4.5 cm \tilde{\epsilon}_2({\bf k})/t = -Z/t+a (k'_y)^2 -
b (k'_x)^2,
\label{16}
\end{equation}
where the vawevectors $k_x$, $k_y$ are defined
as the deviations from the SP wavevector $(0,\pi)$
and the vawevectors $k'_x$, $k'_y$ 
as the deviations from the SP wavevector  $(\pi,0)$.

{\it The divergences (\ref{12}), (\ref{13}) have an
 "excitonic" nature \cite{excitonic} and this is
  the second aspect
of the criticality of the considered QCP}.
What we mean as the "excitonic" nature is that the chemical
potential lies on the bottom of one "band" (around one SP)
and on the top of the another (around another SP) for the
given directions $(0,\pi)-(\pi,\pi)$ and $(\pi,0)-(0,0)$, see Fig.2.
 Therefore, no energy is needed
to excite the electron-hole pair.

\vspace*{-20 mm}
\begin{figure}
\epsfig{%
file=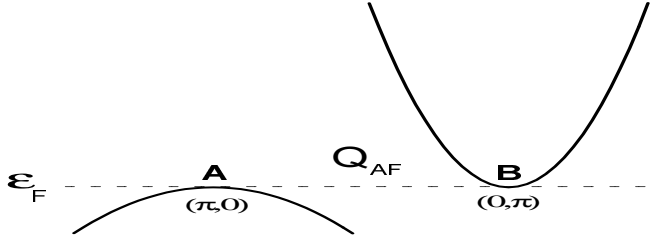,%
figure=figure2.eps,%
height=10cm,%
width=12cm,%
angle=0,%
}\\
\vspace*{-15 mm}
\caption{ Schematical
presentation of the electron
spectrum  in a vicinity of two SP's for $Z=0$.}
\label{f2}
\end{figure}

3. {\it The third aspect of the criticality is related
 to the fact that this point is the end
of a critical line of the static  square-root Kohn singularities in 
$\chi^0({\bf q},\omega)$}. It is this aspect which
gives rise to asymmetrical behaviour of the
system on two sides of QCP, $Z<0$ and  $Z>0$ and
to other anomalies. Below we will concentrate on this aspect
of criticality as has never  been considered before.

\vskip 2 cm

\subsection{The line of static Kohn singularities
 in the vicinity of
${\bf q}={\bf Q}_{AF}$ for $T=0$, $Z<0$}

One can show by simple calculations
that for each point of demiaxis $Z<0$ there
 is a vavevector  ${\bf q = q}_m$ in a
 vicinity of ${\bf Q}_{AF}$
(in each direction around ${\bf Q}_{AF}$) such that

\begin{equation}
Re\chi^0({\bf q}_{},0)-Re\chi^0({\bf q}_{m},0)=
 \left\{
\begin{array}{cc}
A \sqrt{|{\bf q}_{m}-{\bf q}|},
& \hskip 0.3 cm {\bf q}<{\bf q}_{m},\\
B |{\bf q}-{\bf q}_{m}|,
& \hskip 0.3 cm {\bf q}>{\bf q}_{m}.
\end{array}
\right.
\label{17}
\end{equation}
This is a {\bf static}  Kohn
singularitiy in the 2D electron system. The absolute value of
this wavevector in the given direction $\phi$,
$ Q_m(\phi) =|{\bf Q}_{AF}-{\bf q}_m(\phi)|$, is proportional
to the  square-root of the energy distance from QCP :

\begin{equation}
 Q_m(\phi) \propto \sqrt{\mid Z \mid}.
\label{18}
\end{equation}
Therefore  for each fixed negative $Z$ 
 one has a closed line
of the static  Kohn
singularities around ${\bf q= Q}_{AF}$. With
decreasing $|Z|$ the  close line shrinks
and is ended at $Z=0$ being reduced to the point
${\bf q= Q}_{AF}$.
The closed line of the static  Kohn
singularities does not grow again on the other
side of QCP, $Z>0$.

To illustrate this we show in Fig.3 the  wavevector
 dependences of $\chi^{0}({\bf k},0)$ calculated
 based on (\ref{11}) and (\ref{1}). In these plots one sees 
 only a quarter of the
picture around ${\bf q= Q}_{AF}$; to see the {\bf closed}
 line of  square shape around
${\bf Q}_{AF}$ one has to consider the extended BZ around
$(\pi,\pi)$.


\begin{figure}
\epsfig{%
file=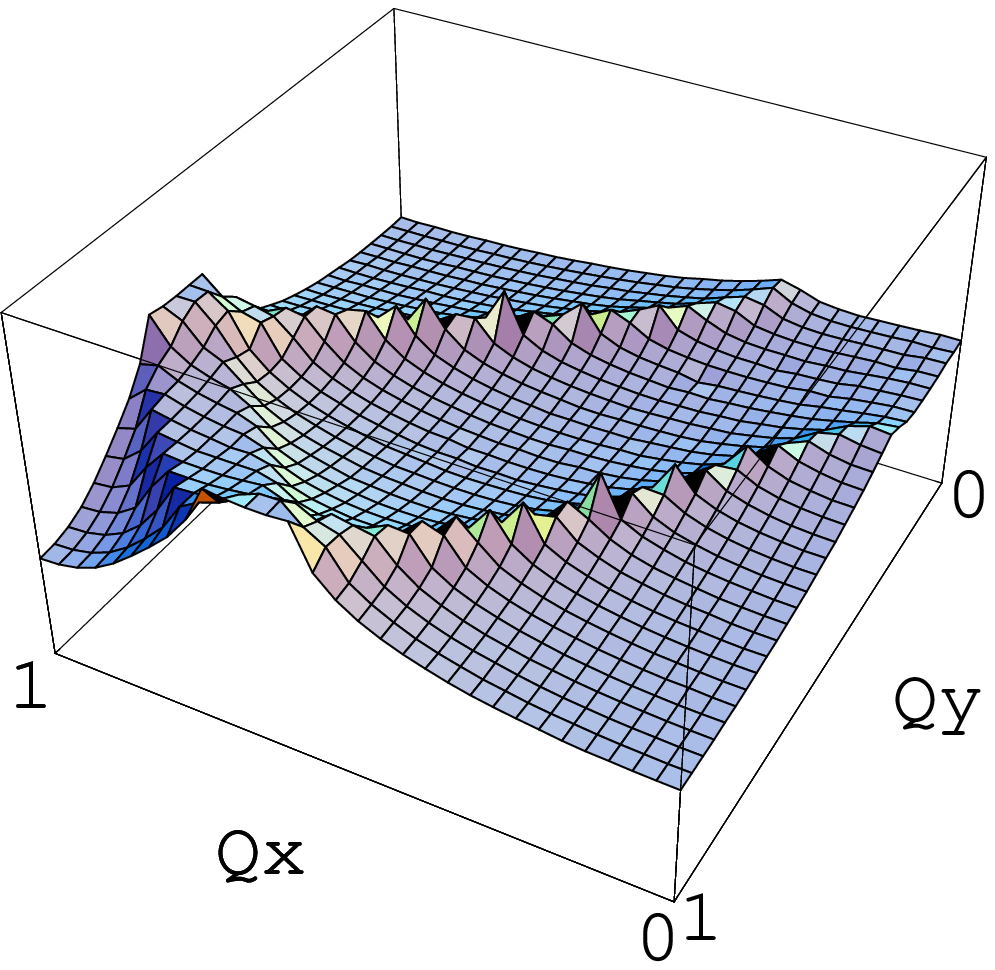,%
figure=figure3.eps,%
height=5cm,%
width=5cm,%
angle=0,%
}
\epsfig{%
file=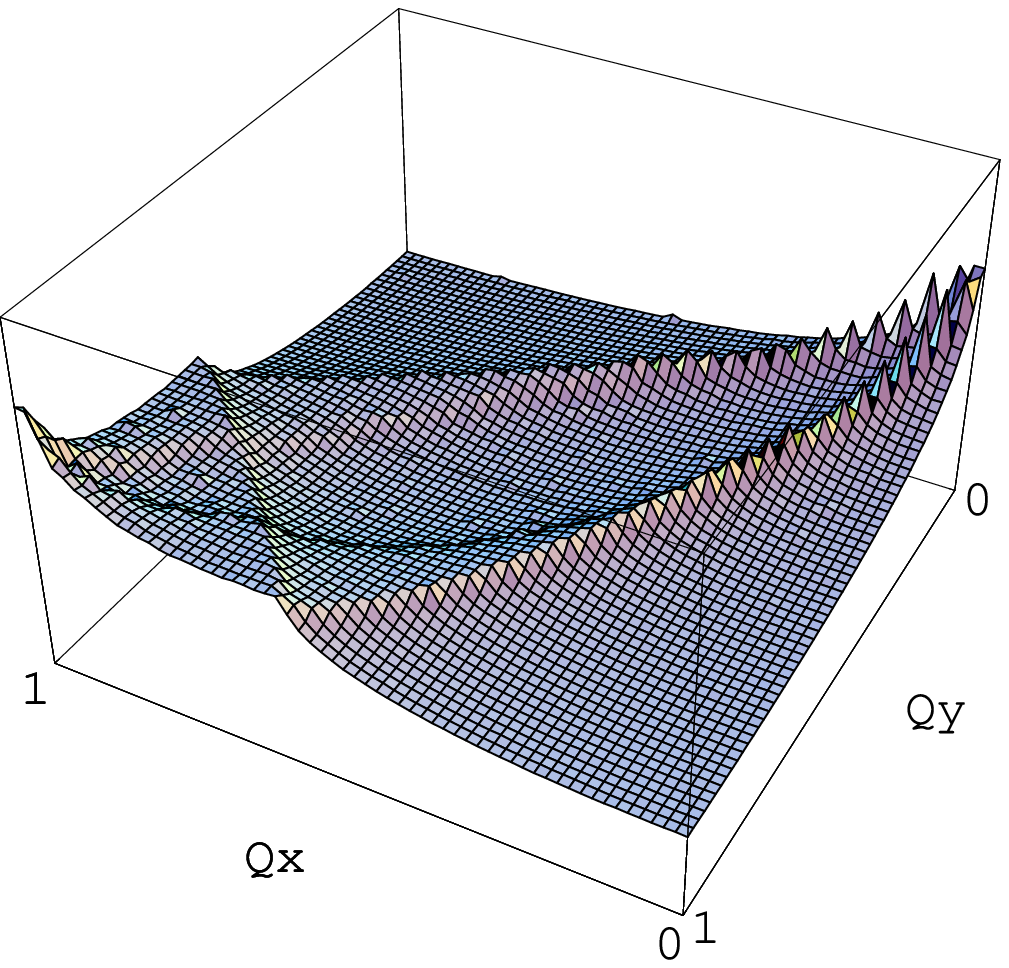,%
figure=figure4.eps,%
height=5cm,%
width=5cm,%
angle=0,%
}
\epsfig{%
file=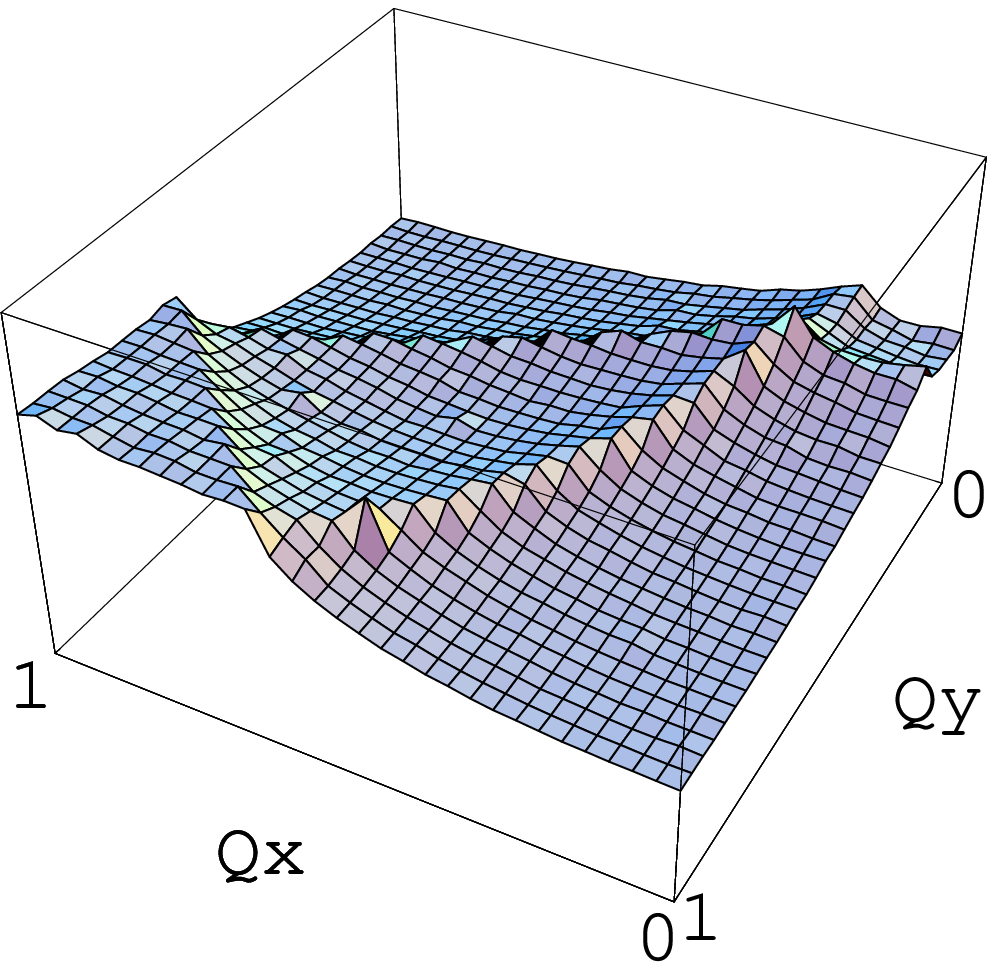,%
figure=figure5.eps,%
height=5cm,%
width=5cm,%
angle=0,%
}
\\
\hspace*{2cm}a)\hspace*{5cm}
b)\hspace*{5cm}c)\\
\caption{Wavevector dependences of the static
 susceptibility $\chi^{0}({\bf q},0)$ for (a) $Z<0$,
 (b) $Z=0$ and  (c) $Z>0$ ($t'/t=-0.3$).
 Here $Q_x=q_x/\pi$,  $Q_y=q_y/\pi$. The point
${\bf q=Q}$ corresponds to the left corner.}
\label{f3}
\end{figure}
The discussed above line is the closest to 
${\bf Q}_{AF}$ line of singularities in Fig.3a. [There are few
other lines of the Kohn singularities seen in Fig.3
which are not sensitive to ETT, 
 we  discuss them in \cite{OPQCP2}] One
can see that this line disappears at $Z=0$ and
does not reappear at $Z>0$ (while the other lines of
Kohn singularities in a proximity of
${\bf Q}_{AF}$
do not change across QCP).

 The wavevector dependence of the static
susceptibility in
the regime  $Z>0$  has a weak maximum
at ${\bf q= Q}_{AF}$ for very small values of $Z$,
for higher $Z$ it exhibits a wide  
  plateau until some border wavevector
  whose value for a given direction depends
on $Z$ only slightly, see Fig.4. 
 To illustrate an  evolution with  $Z$ of the
$\bf q$ dependence of $\chi^{0}({\bf q},0)$
 for both regimes  we present
in  Fig.4  calculations
of  $\chi^{0}({\bf q},0)$
  in the fixed direction, here  $(q_x,\pi)$

\begin{figure}
\begin{center}
\epsfig{%
file=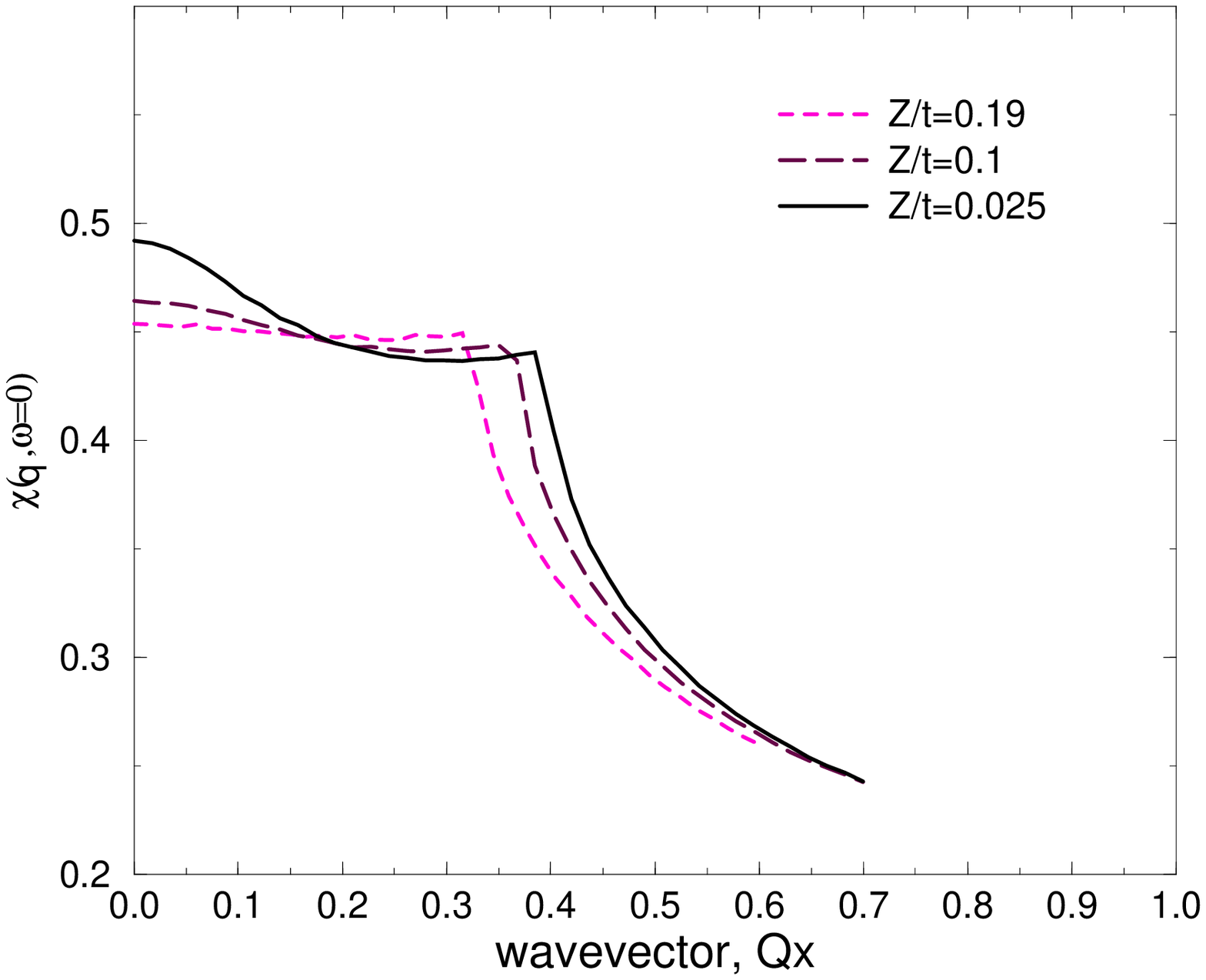,%
figure=figure6.eps,%
height=8cm,%
width=8cm,%
angle=0,%
}
\epsfig{%
file=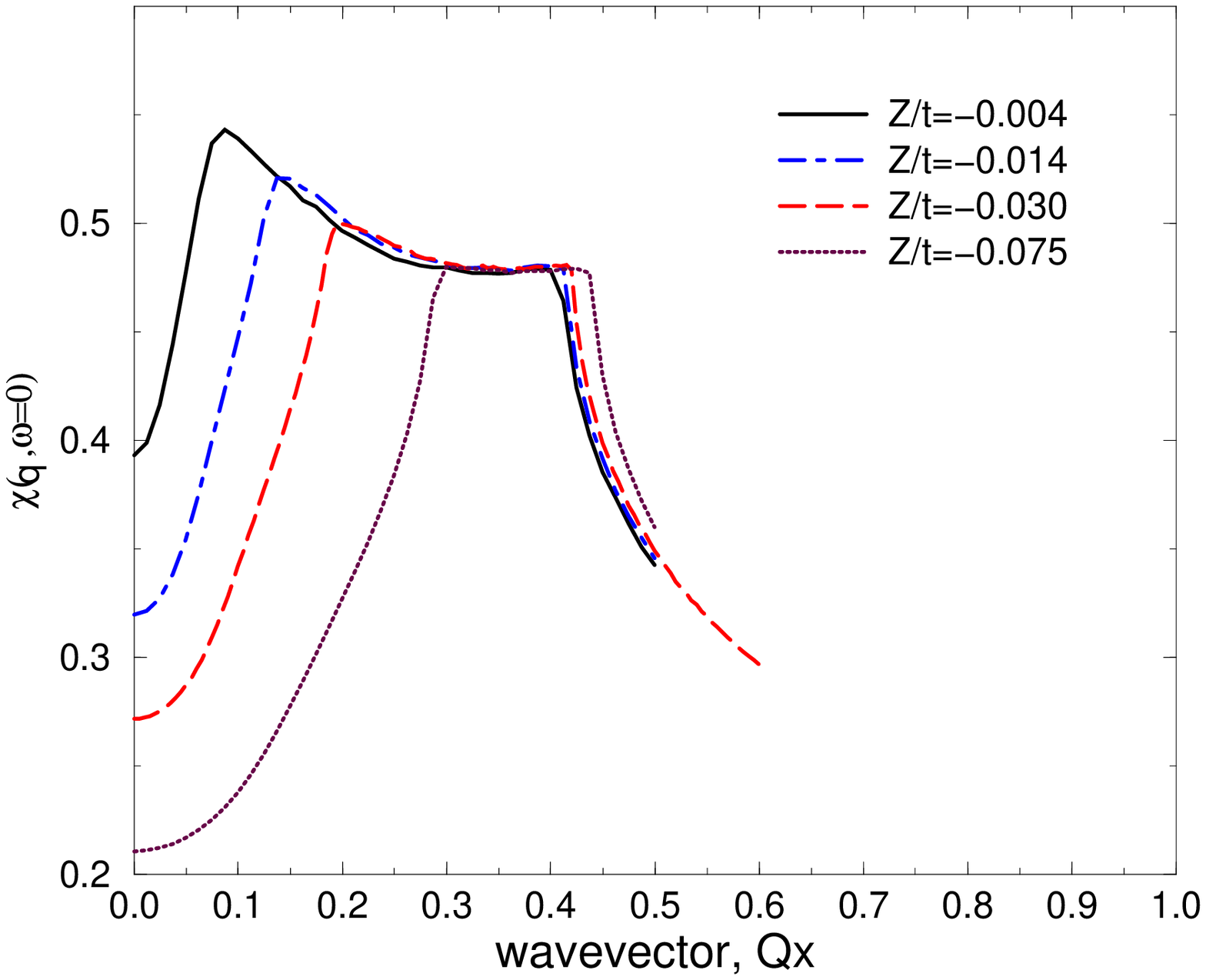,%
figure=figure7.eps,%
height=8cm,%
width=8cm,%
angle=0,%
}
\\
\hspace*{1cm}a)\hspace*{6cm}
b)\\
\end{center}
\caption{Static
susceptibility $\chi^{0}({\bf q},0)$ as a function
of wavevector in the direction $(q_x,\pi)$ for
different values of $Z$ in the cases   $Z>0$ (a) and
 $Z<0$ (b). $t'/t=-0.3$). Here $Q_x=(\pi-q_x)/\pi$.}
\label{f4}
\end{figure}


The difference between behaviour on two sides of
QCP is clear : there is incommensurability
(singularity) on one side and commensurability
on the other. In both cases there is a singularity at
rather high wavevector $Q_x$ whose origin we
discuss in \cite{OPQCP2}.

It is important to know for the following
applications how the plateau  in the regime
$Z>0$ 
evolves with $T$. For this  we show in
Fig.5  the
$\bf q$ dependence of $Re\chi^0$ calculated for fixed $Z$
and increasing $T$.

\begin{figure}
\begin{center}
\epsfig{%
file=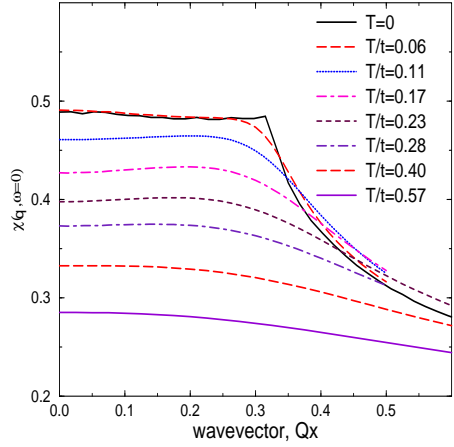,%
figure=figure8.eps,%
height=7cm,%
width=7cm,%
angle=0,%
}
\\
\end{center}
\caption{Static electron-hole susceptibility $\chi_0({\bf q},0)$
as a function of wavevector in the direction $(q_x,\pi)$ for
$Z/t=0.19$ and different temperature ($t'/t=-0.3$)}
\label{f5}
\end{figure}
One can see that the plateau 
survives until very high temperature. 

Above we analyzed the $\bf q$ dependence of the
electron-hole susceptibility  on the critical line $T=0$,
$Z<0$ ($\omega=0$). It is also worth to know the type of
singularities in $T$ and $\omega$ 
for the  susceptibility $\chi^0({\bf q}_{m},Z,\omega,T)$
when approaching the  critical line. They are given by :

\begin{equation}
\chi^0({\bf q}_{m},Z,\omega,T)-\chi^0({\bf q}_{m},Z,\omega=0,
T=0)
\propto
 \sqrt{T-i\omega},  \hskip 1 cm T \rightarrow 0, \omega \rightarrow 0
\label{19}
\end{equation}
for finite $Z$ ($Z>T$, $Z>\omega$).
When approaching the  end of the critical line which 
is the ETT quantum critical point
$T=0$, $Z=0$ ($\omega=0$), the behaviour changes. Now
it depends on the order $Z \rightarrow 0$ and $T \rightarrow 0$, $\omega \rightarrow 0$. When first $T \rightarrow 0$, $\omega \rightarrow 0$
we still have eq. (\ref{19}). When first  $Z \rightarrow 0$ one has

\begin{equation}
\chi^0({\bf Q}_{AF},Z,\omega,T) \propto Ln(T+i\omega).
\label{19a}
\end{equation}

It was the static Kohn singularities which exist only 
on one side of QCP,
$Z<0$, and take place for the incommensurate wavevector
${\bf q} ={\bf q}_{m}$. Below we show 
that on the other
side, $Z>0$, static singularities related to ETT do not exist but
a dynamic Kohn singularity also related to ETT
appears at  ${\bf q} ={\bf Q}_{AF}$.

\subsection{The line of dynamic Kohn singularity for $T=0$, $Z>0$}

Let's analyse the energy dependence of 
$\chi^0({\bf q},\omega)$ in the regime $Z>0$ for
the characteristic wavevector in this regime,
 ${\bf q}={\bf Q}_{AF}$. The calculated dependence
(based on (\ref{11})
 and the full
spectrum (\ref{1})) is shown in Fig.6.
One can see that for fixed $Z$ there is a
 characteristic energy,
 $\omega_c$, where  $Im\chi^{0}$ and
$Re\chi^{0}$ are singular. This energy increases with
increasing $Z$.
There is
a plateau in $Re\chi^{0}$
for $\omega<\omega_c$
and a sharp decrease for $\omega>\omega_c$.
As this behaviour has very important consequences,  we are
going to analyse it analytically.

\begin{figure}
\begin{center}
\epsfig{%
file=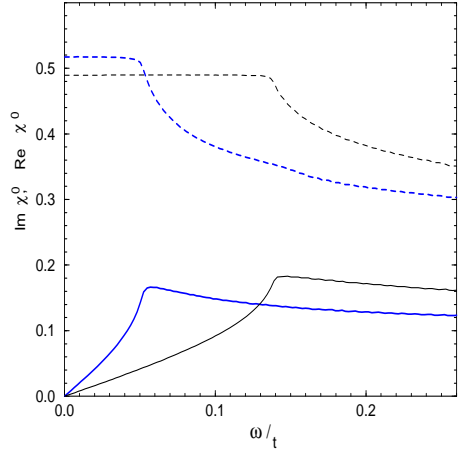,%
figure=figure9.eps,%
height=7cm,%
width=7cm,%
angle=0,%
}\\
\end{center}
\caption{ Energy dependences of $Re\chi^0$ (dashed lines) and
 $Im\chi^0$ (solid lines)
for the wavevector ${\bf q}={\bf Q}_{AF}$.
The thick lines correspond to $Z/t=0.04$, the
thin lines to  $Z/t=0.1$ ($t/t'=-0.3$).}
\label{f6}
\end{figure}

Calculations of $Im\chi^{0}({\bf k},\omega)$
for ${\bf k}={\bf Q}_{AF}$ performed with the hyperbolic
spectrum (\ref{16}) give the scaling expression

\begin{equation} 
Im\chi^{0}({\bf Q}_{AF},\omega)=\frac{1}{2\pi t}
 F(\frac{\omega}{\omega_c},\frac{b}{a}),
\label{20}
\end{equation}
where the scaling function $F(x,y)$ is determined as follows 

\begin{equation}
F(x,y) = 
 \left\{
\begin{array}{cc}
\ln{\frac{\sqrt{1+xy}+\sqrt{1+x}}{\sqrt{1-xy}+\sqrt{1-x}}},   
&  0 \leq x \leq 1\\
\ln{\frac{\sqrt{1+xy}+\sqrt{1+x}}{\sqrt{x(1-y)}}}, &  x \geq 1.
\end{array}
\right.
\label{21}
\end{equation}
 In (\ref{20}), $a$ and $b$ are the coefficients of the dispersion
law (\ref{16}), $b/a$ is equal
$$
b/a=1+\frac{4t'/t}{1-2t'/t}.
$$
and $\omega_c$ is given by

\begin{equation}
 \omega_c=\frac{2Z}{1-2t'/t}.
\label{22}
\end{equation}
[The expression (\ref{21}) is valid for $t' \neq 0$].

Using Kramers-Kronig relation one obtains
the
following equations for $ Re\chi^0$ :

\begin{equation}
t Re\chi^0({\bf Q}_{AF},\omega)=t Re\chi^0({\bf Q}_{AF},\omega_c)
-\Phi(\frac{\omega}{\omega_c},\frac{b}{a}),
\label{23}
\end{equation}
$$
t Re\chi^0({\bf Q}_{AF},\omega_c)=-\alpha\ln{(\omega_c/t)}+\beta.
$$
In (\ref{23}) $\alpha$ is given by

\begin{equation}
\alpha=F(\infty,b/a)/\pi^2= \frac{1}{\pi^2} \ln{(\frac{1+\sqrt{1-4(t'/t)^2}}
{1-\sqrt{1-4(t'/t)^2}})}
\label{24} 
\end{equation}
and the asymptotic form of $\Phi(x,y)$ as a function of $x$
for fixed $y$ is given by :

\begin{equation}
\Phi(x,y) = 
 \left\{
\begin{array}{cc}
\gamma_1(1-x^2), &  x -1 <0\\
\gamma_2\sqrt{x-1},
&  0 \leq x-1 \ll 1\\
\alpha [\ln{\frac{(x-1)t}{\omega_{max}}}], &  x -1 >x^{*}
\end{array}
\right.
\label{25}
\end{equation}
The eq. (\ref{23})-(\ref{25}) are valid for $Z/t<10^{-1}$.
$\omega_{max}$ is a cutoff energy.
The coefficients  $\beta, \gamma_1$ and $\gamma_2$ depend on $t'/t$.
The analytical expressions are given in \cite{Kisselev}.
For example for $t'/t=-0.3$ they are equal to : $\beta=0.218$,
 $\gamma_1 \approx 10^{-2}$, $\gamma_2=0.18$.
 The value of $x^{*}$
is given by $x^{*} \approx 10$.
{\bf The important features} of $Re\chi^0$ are:
{\it  (i) the almost perfect plateau for
$\omega<\omega_c$ since  $\gamma_1$ is extremely small
(that is true for all finite $t'/t$), (ii)
the square-root singularity at
 $\omega=\omega_c$; 
(iii) the logarithmic behavior for large $\omega$ :
$\omega>\omega^{*} \approx 10 \omega_c$
(quantum
critical behaviour)}.

{\bf Thus, we realize that the same root-square singularity
 in $Re\chi^0$ which exists in the regime $Z<0$ for
 ${\bf q}={\bf q}_m$ as $\omega \rightarrow
0^+$ (see eq.(\ref{19})) 
occurs in the regime $Z>0$ for
 ${\bf q}={\bf Q}_{AF}$ as $\omega \rightarrow
\omega_c +0^+$}. 
It is a dynamic 2D Kohn
singularity.
[We consider here only the case
 ${\bf q}={\bf Q}_{AF}$  since it is the wavevector where 
$\chi^0({\bf q},0)$ is maximum and which therefore determines
all properties related to a long-range and short-range ordering.
Dynamic Kohn
singularities for  ${\bf q} \neq {\bf Q}_{AF}$ will be discussed
elsewhere]

It is worth to see how
the
dynamic Kohn singularities appear in  3D
plot of  $\chi_0({\bf Q}_{AF},\omega)$ as a function
of $Z$ and of $\omega$. We show this in Fig.7a.
If one plots the lines of Kohn singularities which end
at QCP in $\omega-Z$ plane one gets a picture shown in
Fig.7b which demonstrates the third aspect of the criticality
of the considered QCP at $T=0$, $Z=0$.

\begin{figure}
\begin{center}
\epsfig{%
file=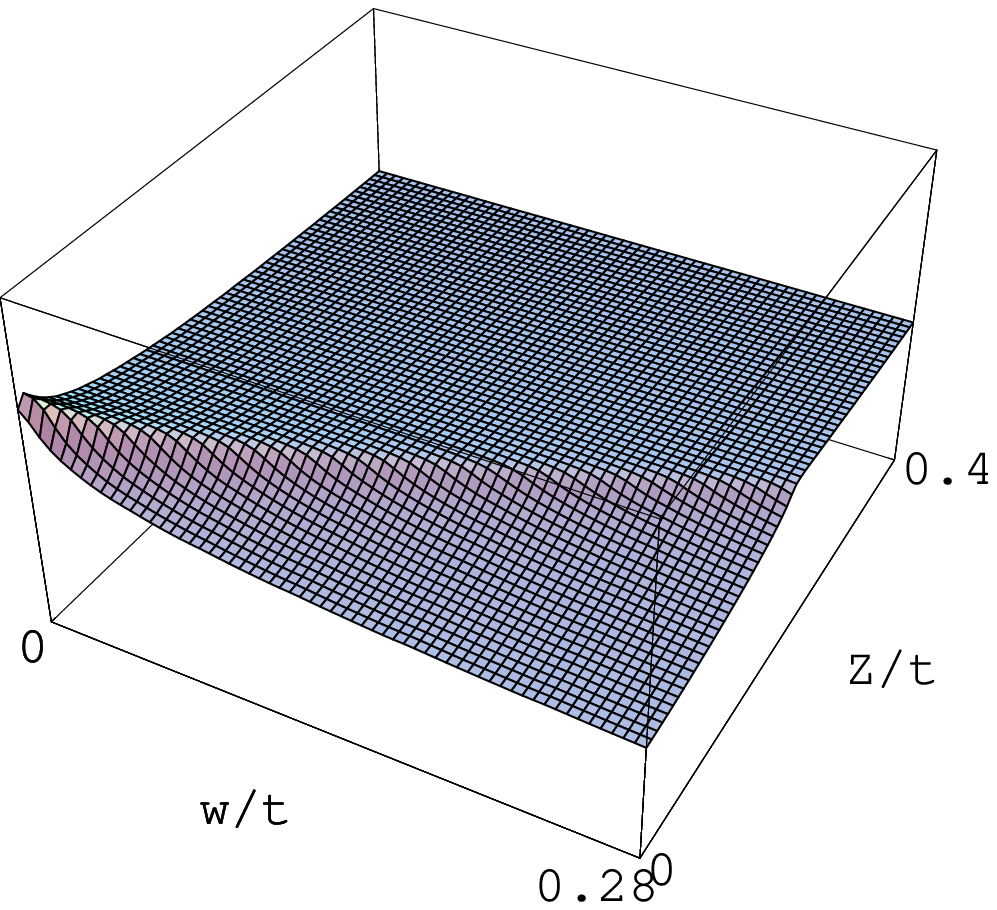,%
figure=figure10.eps,%
height=6.5cm,%
width=6.5cm,%
angle=0,%
}
\epsfig{%
file=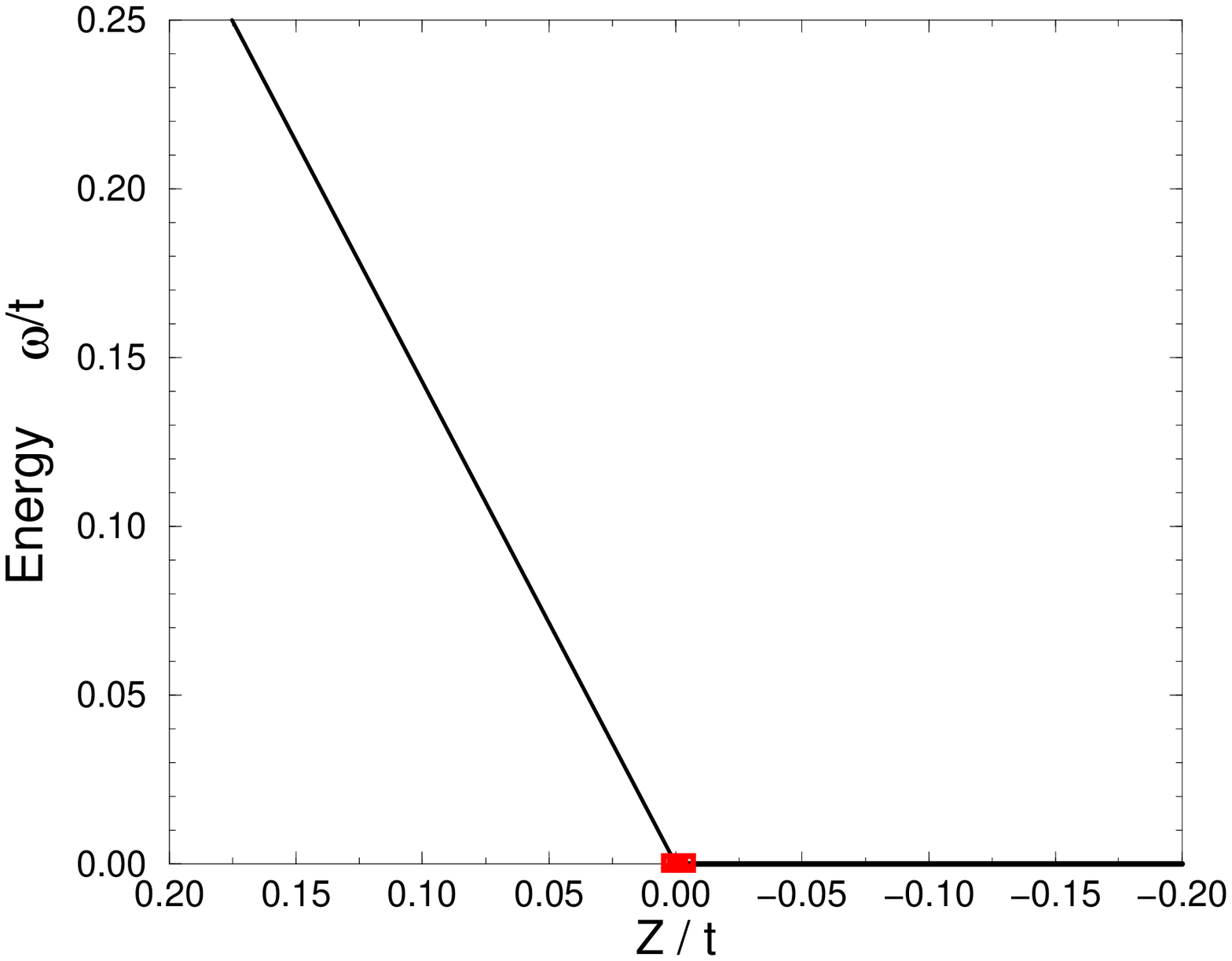,%
figure=figure11.eps,%
height=6.5cm,%
width=6.5cm,%
angle=0,%
}
\\
\hskip 0 cm (a) \hskip 6 cm (b)\\
\end{center}
\caption{ The plots which demonstrate
the third aspect of the criticality of QCP.
(a)  $Re\chi_0({\bf Q}_{AF},\omega)$ as a function
of $Z$ and of $\omega$ in the regime $Z>0$, (b)
 two critical lines of Kohn singularities  :
$\omega=\omega_c^-=0$ for $Z<0$ and $\omega=\omega_c^+=
\omega_c(Z)$ for $Z>0$. ($T=0$, $t'/t=-0.2$)}
\label{f7}
\end{figure}

In fact all properties related to
$\chi_0({\bf Q}_{AF},\omega)$ are singular at
the line $\omega=\omega_c(Z)$ not only
 $Re\chi_0({\bf Q}_{AF},0)$.
In Fig.8 we show as an example the behaviour of
another characteristics 
 $C(\omega)=Im\chi^0({\bf Q}_{AF},\omega)/\omega$
as a function of $\omega$. [This characteristics we will use in
the following sections to analyse NMR experimental data.]

\begin{figure}
\begin{center}
\epsfig{%
file=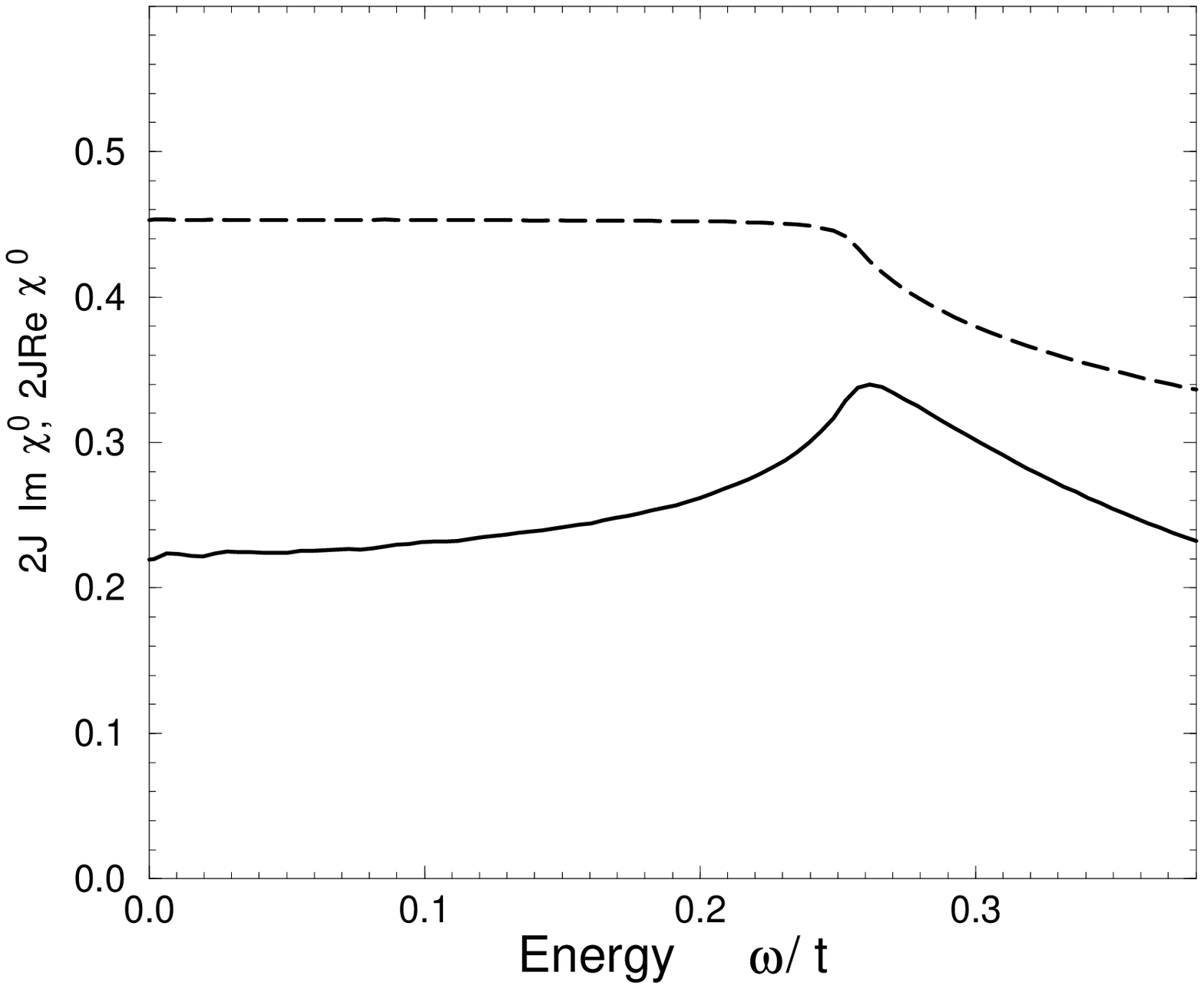,%
figure=figure12.eps,%
height=6.5cm,%
width=6.5cm,%
angle=0,%
}
\\
\end{center}
\caption{ Calculated  $C(\omega)=Im\chi^0({\bf Q}_{AF},\omega)/\omega$ 
(solid line) 
as a function of $\omega$ for the regime $Z>0$ ($t'/t=-0.3$,
$Z/t=0.19$). For comparison we show also
  $Re\chi^0({\bf Q}_{AF},0)$ (dashed line)}
\label{f8}
\end{figure}
One can see that a singularity takes place at 
the same energy as for
$Re\chi_0({\bf Q}_{AF},0)$, i.e. at
$\omega=\omega_c(Z)$.
One can see also that $C$
is constant only at low energies, $\omega \ll \omega_c$
(Fermi-liquid behaviour). At higher energies it
increases with $\omega$ until $\omega=\omega_c$,
passes through the maximum at $\omega=\omega_c$ and
then decreases. It is interesting to emphasize that except
of very low $\omega$ such a behaviour gives an
impression of the existence of a pseudogap in the one-electron
spectrum although the pseudogap is absent in the
bare electron spectrum.

\subsection{The lines of temperature 
"Kohn anomalies" for $Z>0$}

Let's consider now the regime $Z>0$
for finite $T$, namely let's consider
a behaviour of 
$Re\chi^0({\bf Q}_{AF},0)$ as a function of $T$.
Such dependences calculated for two different values of $Z$ 
are presented in Fig.9
 
\begin{figure}
\begin{center}
\epsfig{%
file=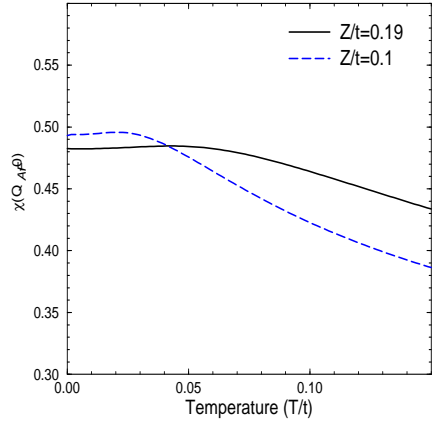,%
figure=figure13.eps,%
height=6.6cm,%
width=6.5cm,%
angle=0,%
}
\\
\end{center}
\caption{ Static electron-hole  susceptibility 
 $\chi_0({\bf Q}_{AF},0)$
 as a function of $T$
 for two values of $Z$ in the regime $Z>0$ ($t'/t=-0.3$).}
\label{f9}
\end{figure}

 When comparing with Fig.6
one can see that the behaviour is almost the same 
being of course smoothed by the effect of finite $T$ :
 there is   a plateau until some temperature $T^*_{Re}$
and a rather sharp decrease at higher temperature.
In the same way as it was for the
characteristic energy, the characteristic temperature
 scales with
$Z$ :

\begin{equation}
T^{*}_{Re} \propto 
Z.
\label{26}
\end{equation}

In fact the behaviour of $Re\chi^0({\bf Q}_{AF},0)$
is slightly different from the plateau behaviour
in the range $T<T^*_{Re}$. There is a
slight maximum at the end of the "plateau". It is
almost invisible in the graph but is quite important and
intrinsic property : {\bf this maximum clearly distinguishes
the point $T^*_{Re}$ which is the point of the temperature
"Kohn anomaly" and therefore the line $T^*_{Re}(Z)$
is the line of Kohn anomalies for $Re\chi^0$.}

 There is a much more pronounced maximum in $Z$ dependence of  $Re\chi^0({\bf Q}_{AF},0)$ at finite $T$ and $Z>0$,
see Fig.10. The position of the maximum is proportional to $T$,
 $Z^*_{Re} \propto T$. The point $Z^*_{Re}$
is the point of the electron concentration Kohn anomaly \cite{scaling}.

Another feature important for following applications is that
 $\chi_0({\bf Q}_{AF},0)$ as a function of $Z$
is assymetrical on two sides from
$Z=Z^*$. On the side $Z>Z^*$ it depends on $Z$ very weakly 
and is {\bf practically constant starting from some threshold
value of $Z-Z^*(T)$} (that is very unusual)
 while for $Z<Z^*$ it always decreases
with increasing $|Z-Z^*(T)|$. The reason for the former behaviour is
discussed in \cite{OPQCP2}.

\begin{figure}
\begin{center}
\epsfig{%
file=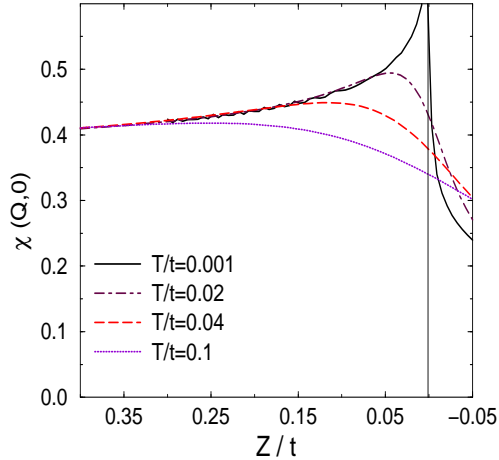,%
figure=figure14.eps,%
height=7cm,%
width=7cm,%
angle=0,%
}
\\
\end{center}
\caption{Static electron-hole  susceptibility 
 $\chi_0({\bf Q}_{AF},0)$ as a function of  $Z$ 
for increasing $T$ ($t'/t=-0.3$). The maximum at some $Z$
which we call $Z^*(T)$ is clearly seen.}
\label{f10}
\end{figure}

It is useful to analyse lines of $Re\chi^0({\bf Q}_{AF},0)=const$
in the $T-Z$ plane
 which we present in Fig.11. These lines have a very unusual form
which reflects the existence of the  Kohn anomalies at $T^*_{Re}(Z)$,
compare for example with the  lines of the ordinary forms
for the SC response function
in Fig.15b. This will have an important
consequence when we will take into account an interaction.

\begin{figure}
\begin{center}
\epsfig{%
file=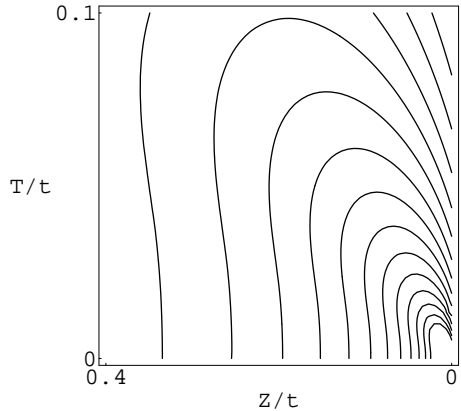,%
figure=figure15.eps,%
height=6cm,%
width=6cm,%
angle=0,%
}
\\
\end{center}
\caption{ The lines of $Re\chi^0({\bf Q}_{AF},0)=const$
in the $T-Z$ plane for the regime $Z>0$.}
\label{11}
\end{figure}

Let's analyse now the temperature dependence of
$C(\omega)=Im\chi^0({\bf Q}_{AF},\omega)/\omega$ calculated
in the limit  $\omega \rightarrow 0$, see Fig.12.
Its behaviour
 repeats in a smooth form  the
behaviour of $C(\omega)$ 
as a function of $\omega$ at $T=0$,
 see Fig.9.
 The important difference is
that {\bf  the characteristic temperatures for
$Re\chi^0({\bf Q}_{AF},0)$ and 
$Im\chi^0({\bf Q}_{AF},\omega)/\omega$
are different} on the contrary to the characteristic energy
at $T=0$ which is the same for both $Re\chi^0$ and
$Im\chi^0$. This is a usual effect of finite temperature.
Of course both characteristic temperatures,
$T^*_{Re}$ and $T^*_{Im}$ are proportional
to $Z$  originating from the same effect of the
Kohn singularities at $T=0$. It is important 
to note that  $T^*_{Im}$ is
always larger than
 $T^*_{Re}$.

\begin{figure}
\begin{center}
\epsfig{%
file=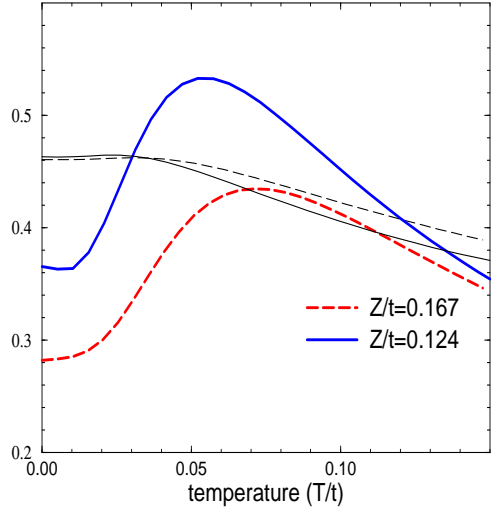,%
figure=figure16.eps,%
height=8cm,%
width=8cm,%
angle=0,%
}
\\
\end{center}
\caption{Temperature dependence of 
 $C(\omega)=Im\chi^0({\bf Q}_{AF},\omega)/\omega$ 
in
the limit $\omega \rightarrow 0$ 
  for two values
of $Z$ in the regime $Z>0$ (thick lines)
($t'/t=-0.3$)
 (the existence of SC state
is ignored). One can see that $C(0)$ remains constant until some
temperature, $T_F$, (Fermi-liquid behaviour), then
increases with $T$ and passes through the maximum at
$T=T^*_{Im}(Z)$. For comparison we show also
 $Re\chi^0({\bf Q},0)$ for the same $Z$ (thin lines) which
exhibits a weak maximum at $T=T^*_{Re}(Z)$.}
\label{f12}
\end{figure}

If we plot the lines of the temperature Kohn anomalies
$T^*_{Re}$ and $T^*_{Im}$
in the $T-Z$ plane we get a picture shown in Fig.13
which will be the reference picture for the system
in the presence of interaction.

\begin{figure}
\begin{center}
\epsfig{%
file=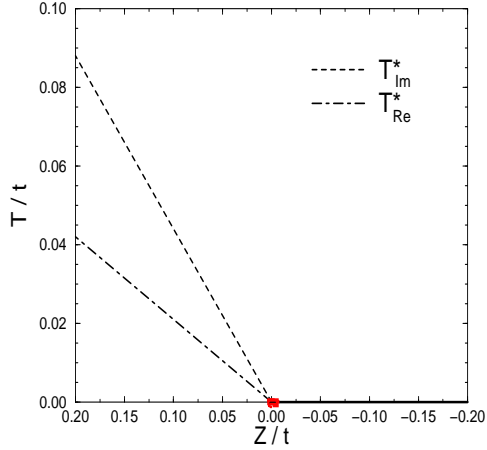,%
figure=figure17.eps,%
height=7cm,%
width=7cm,%
angle=0,%
}
\\
\end{center}
\caption{ The  lines of the 
 temperature Kohn anomalies $T^*_{Re}$  and 
$T^*_{Im}$  ($t'/t=-0.3$). We also show the
critical line $T=0$, $Z<0$ (solid line)}
\label{f13}
\end{figure}

It is worth to give 
one more example of the role of the lines $T^*(Z)$. In
Fig.14a we show a temperature evolution of the $\omega$
dependence of $Re\chi^0({\bf Q}_{AF},\omega)$.
 The low energy plateau survives only until $T=T^*_{Re}$. 
The size of the plateau is :

\begin{equation}
\omega_c(Z,T) = \omega_c(Z,0)- a_1 T
\label{27}
\end{equation} 
($a_1 T^*_{Re}(Z)=\omega_c(Z,0)$).
{\bf Above
$T=T^*_{Re}$ the plateau disappears} and  the behaviour becomes
ordinary. 
In
Fig.14b we show a temperature evolution of the $\omega$
dependence of $C(\omega)$.
The low energy increasing part survives only
 until $T=T^*_{Im}$. 
{\bf Above
$T=T^*_{Im}$, the increasing part disappears}

\begin{figure}
\begin{center}
\epsfig{%
file=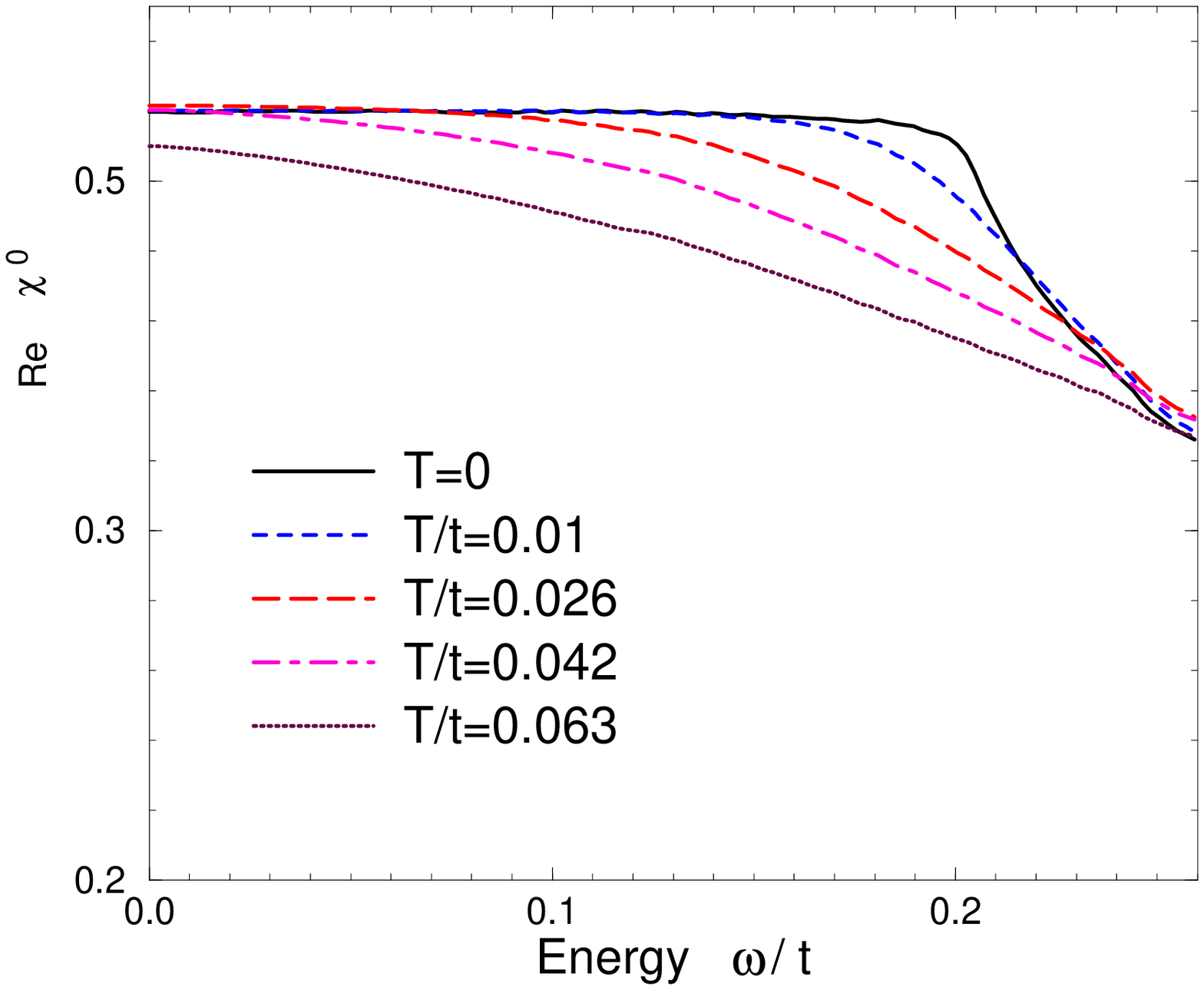,%
figure=figure18.eps,%
height=7cm,%
width=7cm,%
angle=0,%
}
\epsfig{%
file=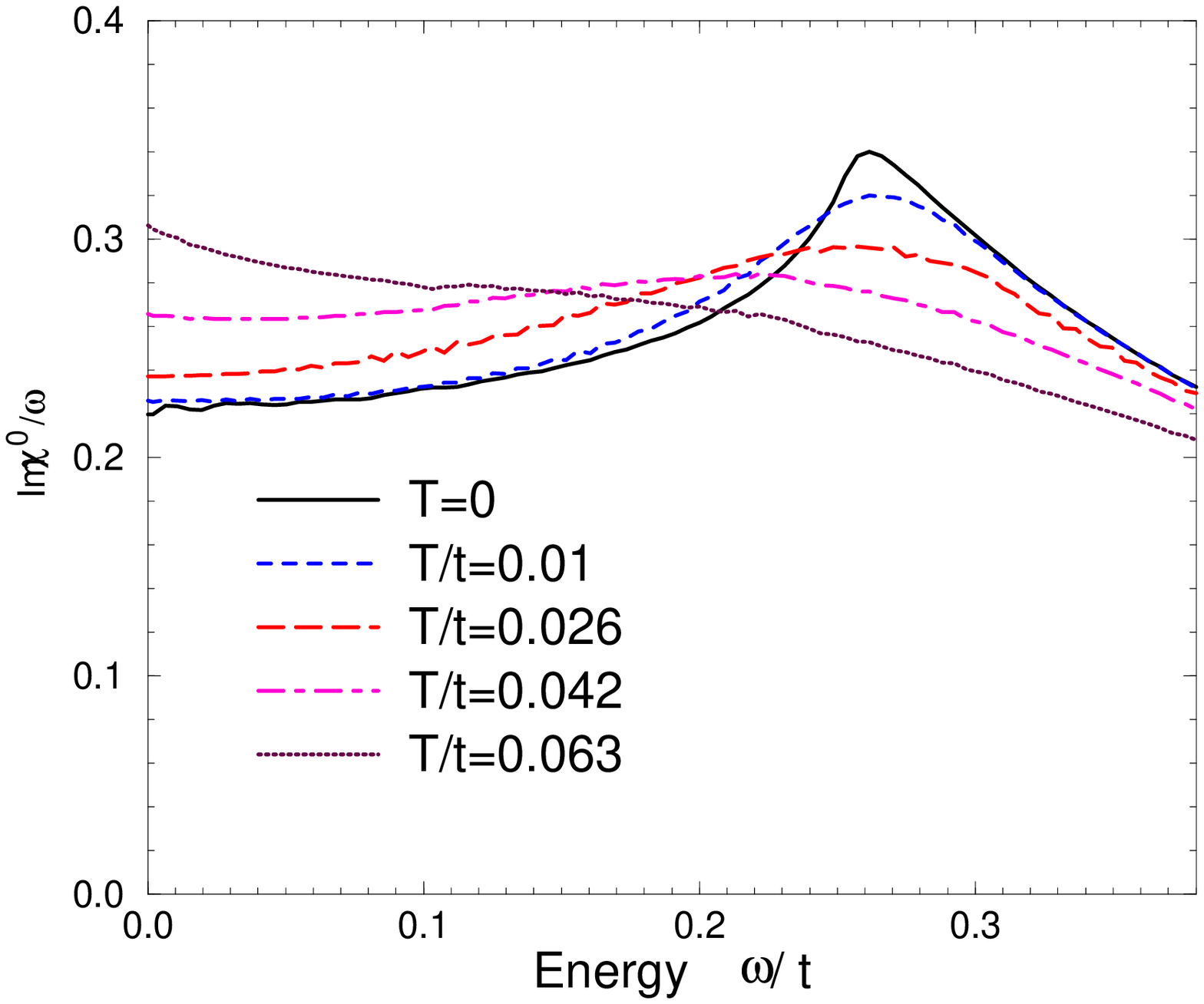,%
figure=figure19.eps,%
height=7cm,%
width=7cm,%
angle=0,%
}
\\
\hskip 1 cm (a) \hskip 5 cm (b)\\
\end{center}
\caption{ Evolution  with temperature (a) of the plateau in $Re\chi^0({\bf Q},0)$ 
 ($Z/t=0.17$,   
$T^*_{Re}/t=0.04$) and (b) 
of the increasing part in $C(\omega)$
($Z/t=0.19$,  
$T^*_{Im}/t=0.05$). ($t'/t=-0.3$)}
\label{f14}
\end{figure}

The described above anomalous behaviour
in the system
of noninteracting electrons leads to
very important consequences when taking into
account an interaction.

\subsection{Superconducting polarization operator
as a function of $Z$ and $T$}

Before we start to consider the system in the presence of
interaction it is worth to analyse the behaviour
of the superconducting polarization operator
$\Pi^0(0,0)$ as a function of $Z$ and of $T$. This
behaviour is presented in Fig.15.

\begin{figure}
\begin{center}
\hskip -1.5 cm
\epsfig{%
file=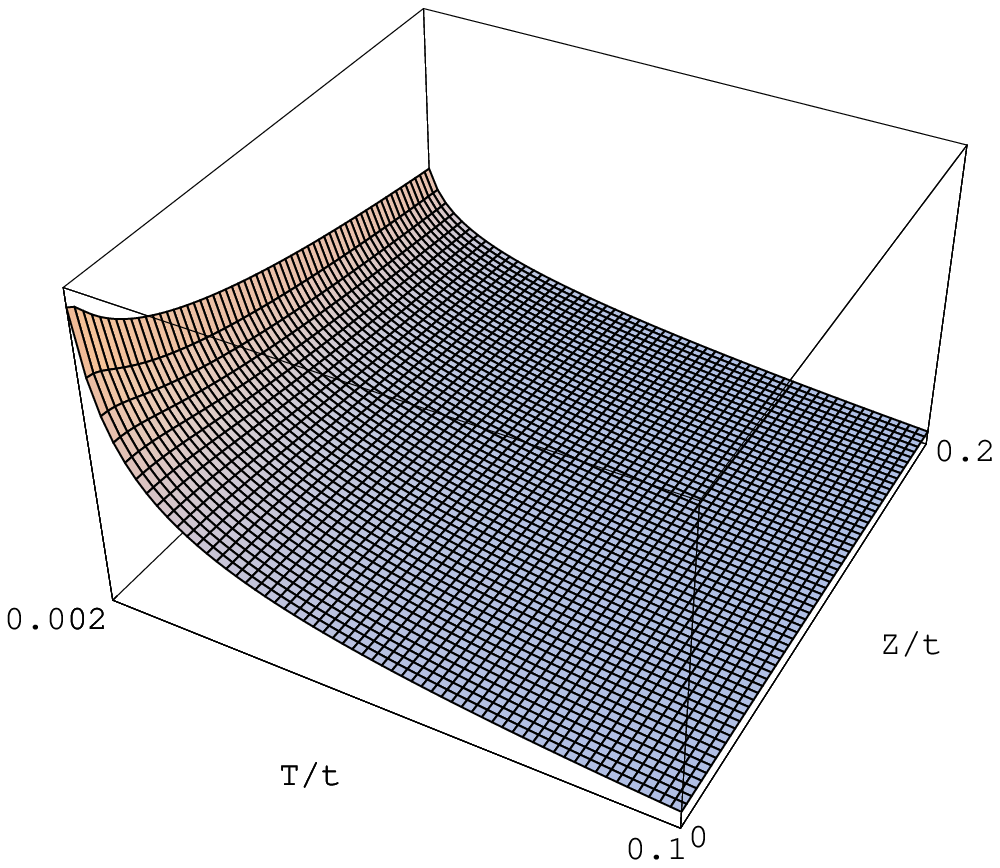,%
figure=figure20.eps,%
height=8cm,%
width=8cm,%
angle=0,%
}
\hskip 1 cm
\epsfig{%
file=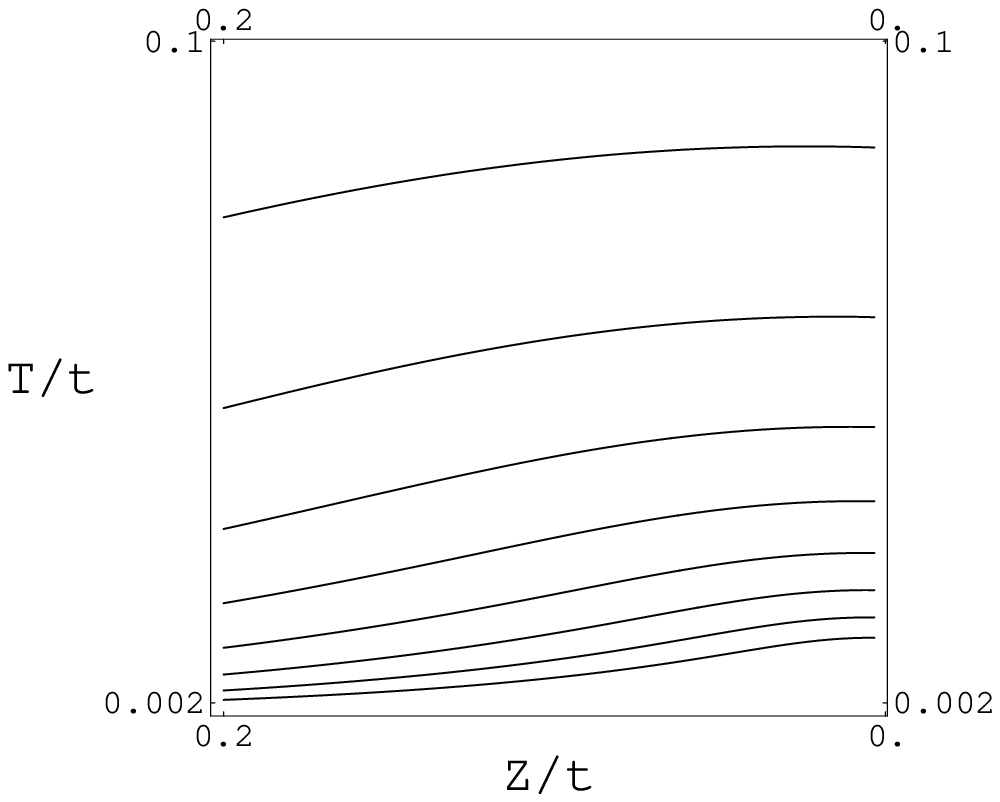,%
figure=figure21.eps,%
height=7.5cm,%
width=7.5cm,%
angle=0,%
}
\\
\hskip 1 cm (a) \hskip 5 cm (b)\\
\end{center}
\caption{ $\Pi^0(0,0)$ as a function of $Z$ and of $T$
(a) and lines of $\Pi^0(0,0)=const$ in $Z-T$ coordinates (b)
for $Z>0$. The behaviour for $Z<0$ is totally symmetrical.
The plot is started from $T=0.002$ since for $T=0$ there is
a logarithmic divergence for any $Z$.
($t'/t=-0.2$)}
\label{f15}
\end{figure}

One can see that it is an ordinary behaviour 
as a function of $T$ for fixed $Z$ and
as a function of $Z$ for fixed $T$ :
$\Pi^0(0,0)$ decreases sharply as a function
of $T$ for fixed $Z$ and  as a function
of $Z$ for fixed $T$. In fact the expressions (\ref{8}), (\ref{9}) work
quite well for finite $T$ and $Z$ contrary to the
case of $\chi^0({\bf Q}_{AF},0)$.
This behaviour reflects
the first aspect of the criticality of the
considered QCP. We also present in Fig.15 the
lines of $\Pi^0(0,0)=const$ in the $T-Z$
plane which have a quite ordinary monotonous form, compare  with 
similar lines for  $\chi^0({\bf Q}_{AF},0)$
 in Fig.11.

 It is worth  to note
that the same qualitatively behaviour takes place for 
the effective "polarization operator" 
${\sf \Pi}^0_d(0,0)$ determined as follows

\begin{equation}
{\sf \Pi}^0_d(0,0)=
\frac{1}{4}\sum_{\bf k}
\frac{(\cos{k_x} - \cos{k_y})^2
tanh(\frac{\tilde{\epsilon}({\bf k})}{2T})}
{ 2\tilde{\epsilon}({\bf k})}.
\label{29}
\end{equation}
It replaces the polarization operator
$\Pi^0(0,0)$
in
the case when the interaction responsible
for the superconductivity is momentum dependent which
leads to the d-wave symmetry of SC order
parameter \cite{OnufrievaPRB96}.

\subsection{A passage from the energy
distance from QCP, $Z$, to the electron concentration}

Above we have considered all properties as functions
of the energy distance from the QCP. It is worth for applications 
to cuprates to change the description and to consider
physical properties as functions of
 electron concentration $n_e$ or of hole doping $\delta=
1-n_e$.
To do such a passage we have to use a relation between $Z$
(or the chemical potential $\mu$) and the hole doping. To get
this relation we will use the condition (for $T=0$) :

\begin{equation}
1-\delta= \frac{1}{N}\sum_{k\sigma}n^{F}(\epsilon_{{\bf k}\sigma}-\mu).
\label{30}
\end{equation}
This condition corresponds to the
electron FS with a volume $V_e \propto 1-\delta$.
Experimentally observed in cuprates "large" FS which exists
in the metallic state
even at quite low doping corresponds to the
condition $V_h \propto 1+\delta$  which is equivalent to eq (\ref{30}).
On the other hand, as known, the undoped materials  
 ($\delta=0$) correspond to a
localized-spin AF state that demonstrates that the
latter condition  is
certainly not correct for very low doping.

The problem of the volume of FS in cuprates,
let say, $\delta$ versus $1+\delta$ (for the
hole FS) is an 
independent problem which we do not want
to discuss in details here. We would like to
note only that we have reached some progress
in the understanding of similar problem for
 the $t-J$ model \cite{OnufrievaPRB96}
for which we have written (based on
the diagrammatic technique for $X$-operators and
using the first approximation in inverse number
of nearest neighbours)
an equation for the chemical potential as a function of
doping \cite{OnufrievaPRB96}. We have shown
that it is similar in a certain sense to
the Van-der-Waals equation : in a certain interval of
doping it has  
two "physical" and one "unphysical" solutions  for each doping.
Among "physical" solutions one corresponds to the state with
localized spins : electrons initially present (at $\delta=0$) are 
localized, FS is formed by
 doped holes only 
 (Curie constant is constant as $T \rightarrow 0$,
 the volume of FS is proportional to
 $\delta$). This solution is unstable against AF
ordering of {\bf localized-spin type}. The second
"physical" solution corresponds to the state
with all holes delocalized : the
 volume of FS is proportional to
 $1+\delta$ while the Curie constant
tends to zero as $T \rightarrow 0$.
In the case $t'/t<0$ at low doping only solution
of the former type exists while at high doping 
only the second one is possible. Postponing 
a quantitative  analysis telling which phase  is favorable
at intermediate doping we presume (as a matter of fact
for cuprates) that it is the second phase which is favorable and
we consider at present only this state
described at $T=0$ by the condition (\ref{30}).

 A value of the critical
doping $\delta_c$ corresponding to the QCP at $Z=0$ 
is determined by the condition (\ref{30}). It depends on 
value of $t'/t$, see Fig.16.

\begin{figure}
\begin{center}
\epsfig{%
file=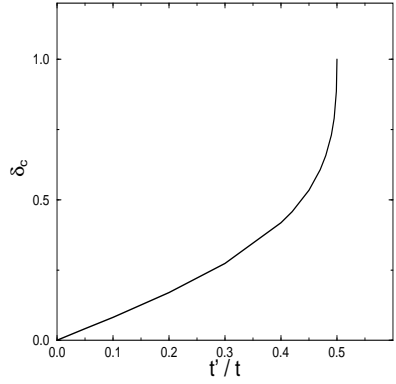,%
figure=figure22.eps,%
height=6cm,%
width=6cm,%
angle=0,%
}
\\
\end{center}
\caption{ Critical doping of ETT $\delta_c$ as a function
of $t'/t$. 
 As realistic values for cuprates  one gets for
example, $\delta_c=0.17$ when $t'/t=-0.2$, $\delta_c=0.27$
when  $t'/t=-0.3$. The value  $t'/t=-0.5$ is critical :
the FS is degenerated into two lines.}
\label{f16}
\end{figure}

So far as
\begin{equation} 
Z \propto \delta_{c}- \delta,
\label{31}
\end{equation}
all dependences considered above can be rewritten
as  functions of doping distance from QCP.

\vskip 2 cm

\section{The system in the presence of interaction}

\subsection{Phase diagram}

In the presence of interaction 
the expression for the electron-hole susceptibility
 in the simplest RPA approximation
is given by 

\begin{equation}
\chi({\bf q},\omega)=
\frac{\chi^0({\bf q}, \omega)}{1+V_{\bf q}\chi^0({\bf q},
 \omega)}.
\label{32}
\end{equation}
The explicit form of the interaction depends on
model. One can consider an interaction which leads
to SDW or CDW instabilities or to both of them.
For example, for
the Hubbard model  $V_{\bf q}=-U$, for the
$t-J$ model  $V_{\bf q}=J_{\bf q}$ ($J_{\bf q}=
2J(\cos q_x+\cos q_y)$). So far as both wavevectors
${\bf q}={\bf Q}_{AF}$
and ${\bf q}=0$ are critical for the considered QCP
and $Re\chi^0({\bf q},0)$ diverges for both of
them whereas experimentally for cuprates only
a response around ${\bf q}={\bf Q}_{AF}$
is observed and it is a spin dependent response
 one should consider this as
a phenomenological argument in a favour of the
momentum dependent interaction in a triplet
channel. From now on we shall mainly consider
 the case of cuprates and therefore 
 we shall use
$V_{\bf q}=J_{\bf q}$ with a positive
sign, $J>0$.

The line of SDW instability 
associated with the considered QCP is given by

\begin{equation}
1/J_{\bf q}=-\chi^0({\bf q}, 0).
\label{33}
\end{equation}
It is clear from the previous analysis that
the instability occurs at ${\bf q}={\bf Q}_{AF}$
on the side $Z>0$ (or $\delta<\delta_c$)
and close to ${\bf q}={\bf q}_{m}$ on the side $Z<0$
(or $\delta>\delta_c$).
Below we will call this instability the {\bf SDW
"excitonic" instability}  in order to distinguish
from the SDW instability associated with nesting
of Fermi surface occuring in the case $t'=0$. In the
latter case all discussed in the paper anomalies originated
from
the dynamic Kohn singularities disappear: the behaviour of $\chi({\bf q},\omega)$
is symmetrical in $Z$ and
corresponds to that in the regime $Z<0$ in 
the case $t' \neq 0$. The reason is discussed in \cite{OPQCP2}.

From  Fig.11  which shows the lines of 
$Re\chi^0({\bf Q}_{AF},0)=const$ 
in $T-Z$ coordinates it is clear that the 
critical line $T_{exc}(Z)$
would have an unusual shape as a function of $Z$
in the regime $Z>0$.
Indeed, we see in Fig.17  that $T_{exc}(Z)$ {\bf increases
with increasing the distance from QCP}
instead of having the form  of
a "bell jar" around QCP as it  usually happens
for an ordered phase developing around an ordinary quantum critical
point and as indeed it occurs on the side $Z<0$ \cite{finiteT}.

\begin{figure}
\begin{center}
\epsfig{%
file=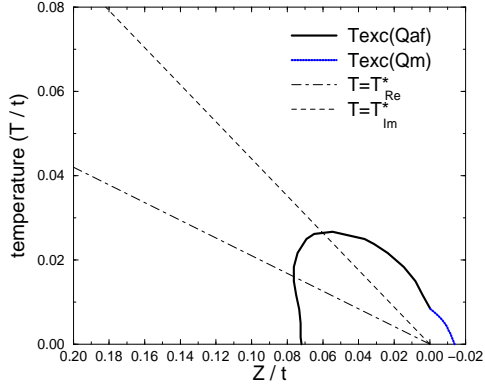,%
figure=figure23.eps,%
height=6cm,%
width=7cm,%
angle=0,%
}\\
\vspace*{-0mm}
\end{center}
\caption{ Phase diagram around QCP, $T=0$, $Z=0$.
 The solid line
corresponds to the SDW "excitonic" instability
 occuring
at ${\bf q} = {\bf Q}_{AF}$ in the  regime,
$Z>0$ and  at
the incommensurate wavevector ${\bf q} = {\bf q}_m$
in the  regime,
$Z<0$.
We show also the 
 lines  $T^*_{Re}(\delta)$ and $T^*_{Im}(\delta)$ discussed in the text.
($t/J=1.9$, $t'/t=-0.3$)}
\label{f17}
\end{figure}

{\bf The form of $T_{exc}(Z)$
reflects the fact that the SDW "excitonic"  ordered
phase develops rather "around" the line $T^*_{Re}(Z)$
than around the point $T=0,Z=0$}.

 On the contrary, the form of
the critical line for the SC phase,
  $T_{sc}(Z)$, which develops around the 
considered QCP
is very ordinary as it repeats the form of the lines in
Fig.15.b
 Whatever is
the nature of the interaction 
responsible for the existence of high $T_c$
 superconductivity,
the line $T_{sc}(Z)$ has the usual shape of
"bell" being symmetrical for the regimes
$Z>0$ and $Z<0$ (or $\delta<\delta_c$ and $\delta>\delta_c$)
which therefore we can call 
{\bf the underdoped and overdoped regimes}, respectively.
The ordinary form is related to the ordinary
 behaviour of $\Pi^0(0,0)$ and $\Pi_d^0(0,0)$
as a function of $Z$ and of $T$ as
discussed in the Sec.II.E. We will not discuss details
 concerning  the SC phase here (see \cite{OnufrievaPRB96} where the line
of SC instability
is discussed from below and
\cite{OPINS} where it
is discussed from above].

Anyway there are two possibilities :
(i) the ordered SDW "excitonic"
 phase leans out OF the SC phase as in Fig.18a,
(ii) the ordered SDW "excitonic"
 phase is completely hidden 
under the SC phase as in Fig.18b.
It depends on the ratio $J/W$, where $W$ is
a bandwidth of the electron band (detailed calculations
for SC critical line are presented
in \cite{OPINS}). [An interplay and a mutual influence of the
two ordered phases, SDW "excitonic"  and SC, will be discusses
elsewhere].

\begin{figure}
\begin{center}
\epsfig{%
file=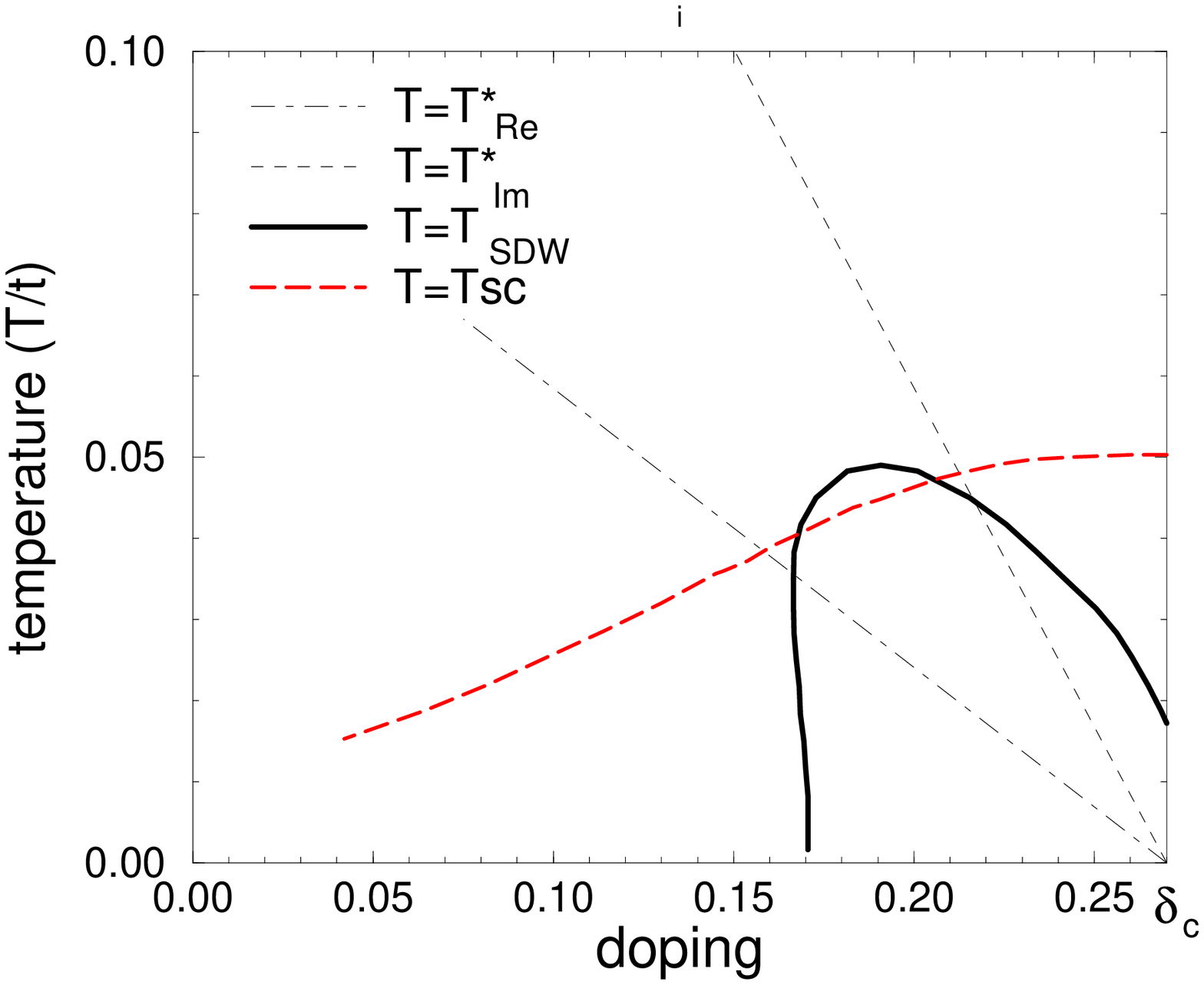,%
figure=figure24.eps,%
height=6cm,%
width=6cm,%
angle=0,%
}
\epsfig{%
file=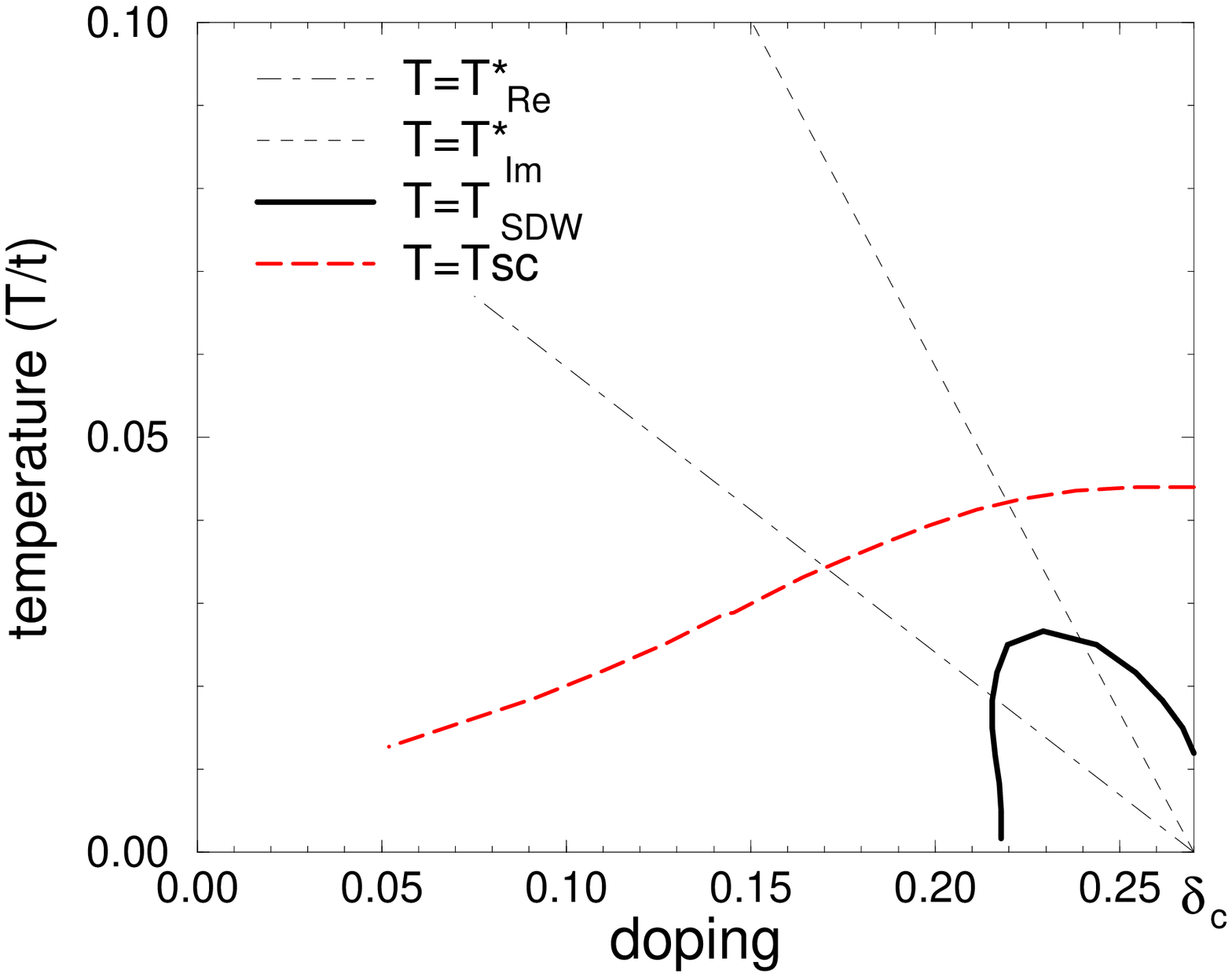,%
figure=figure25.eps,%
height=6cm,%
width=6cm,%
angle=0,%
}
\\
\vspace*{-0mm}
\hskip 1 cm (a) \hskip 5 cm (b)\\
\end{center}
\caption{ Phase diagram in coordinates $T-\delta$
with the lines of SDW
"excitonic" instability , SC instability
 and the  lines  $T^*_{Re}(\delta)$
and  $T^*_{Im}(\delta)$
 discussed in the text ($t'/t=-0.3$, $t/J=1.8$ in (a) and
$t/J=1.9$ in (b)). We show only the
underdoped part of the  phase diagram.} 
\label{f18}
\end{figure}

In both cases  the lines $T^*_{Re}(\delta)$ and
$T^*_{Im}(\delta)$ persist
above  $T_{sc}(\delta)$ in the underdoped regime and therefore
it is worth to analyse properties of the normal state
in the underdoped regime above and below  these lines.

\subsection{ Metallic state in the 
proximity of the ordered SDW "excitonic" phase.}

Below we consider some properties in the undistorted
metallic state out of the ordered SDW "excitonic" phase which
are characterized by anomalous behaviour.

Taking into account the $\omega$ and $\bf q$-dependence of
$Im\chi^0$ and $Re\chi^0$ given by (\ref{20})-(\ref{25}) and the
definition (\ref{32}) one can present  the  electron-hole
susceptibility
 describing fluctuations around ${\bf Q}_{AF}$ in
a proximity of the SDW "excitonic" phase in the form

\begin{equation}
\chi({\bf q},\omega)=\frac{1}{4J}
(\frac{\alpha+iC\omega}{(1-\alpha)+
A({\bf q-Q}_{AF})^2-iC\omega}), \hskip 1 cm
\omega < \omega_c(T)
\label{34}
\end{equation}
with $\alpha$ which is defined as 

\begin{equation}
\alpha=4J \chi^0({\bf Q}_{AF}, 0)
\label{35}
\end{equation}
and which is close to unity in THE proximity of the ordered phase.
 The form (\ref{34})
is valid  for $\bf q$ in the vicinity of ${\bf Q}_{AF}$
(${\bf Q}_{AF}-{\bf q} \ll 1$)  and
 for  $\omega < \omega_c(T)$ where the latter is defined by (\ref{27}).

 The term with $A$ appears only from the $\bf q$
dependence of the interaction since the
$\bf q$
dependence of $Re\chi^0({\bf q},0)$ exhibits the plateau
around  ${\bf Q}_{AF}$ which size
depends very slightly on $Z$ and $T$, see Fig.4a
and Fig.5. Therefore
 the parameter $A$ is constant
as a function of $T$ and $Z$ (or $\delta$)
being proportional to the interaction :

\begin{equation}
A(\delta,T) \propto J/t.
\label{36}
\end{equation}

THE absence in the denominator of (\ref{34}) of a real term containing
$\omega$ is a consequence of the plateau in the $\omega$ 
dependence of  $Re\chi^0({\bf q},0)$
which survives until $\omega=\omega_c(T)$
(see  Fig.14).
It gives a restriction in $\omega$ for the form (\ref{34}) :
for $T=0$ it is valid
for $\omega < \omega_c$ while for higher
temperature,  $T < T^*_{Re}(\delta)$, it is valid
only for
$\omega \ll \omega_c$. Moreover the variant of (\ref{34}) with $C$
 not depending on $\omega$ is correct only for
$\omega \ll \omega_c$  even at low temperature. To enlarge
the range of validity one can replace $C$ by $C(\omega)$
considering it as depending on $\omega$ (a quite untrivial
point).
This dependence at
$T=0$ is given by

\begin{equation}
C(\omega)=\frac{4J}{2\pi t}
\frac{F(\omega/\omega_c)}{\omega},
\label{37}
\end{equation}
where $F(x)$ is determined by (\ref{21}).
The explicit dependence of $C(\omega)$ on 
$\omega$ for the case $T=0$ and for finite $T$ is seen from Fig.14.
One has to note also that at $T=0$

\begin{equation}
C(0) \rightarrow \infty \hskip 0.5 cm  as \hskip 0.5 cm
Z\rightarrow 0.
\label{38}
\end{equation}

{\bf Conclusion }: the form (\ref{34}) is useful only for analysis of
static and quasi static properties \cite{susceptibility}.

Let's introduce   the parameter 

\begin{equation}
\kappa^2_{exc}=1-\alpha
\label{39}
\end{equation}
which as seen from (\ref{34}) is equal to zero on the line of the phase
transition  and determines a proximity  to SDW "excitonic" phase
ordered phase
for the
disordered metallic phase.
The behaviour of $\kappa^2_{exc}$
as a function of
temperature- and doping- distances from the ordered phase
is extremely unusual: {\it (i) $\kappa^2_{exc}$
 is
minimum at the line $T^*_{Re}(\delta)$ not
at $T=0$, (ii) it remains very low in the whole
range of doping and temperature below  $T^*_{Re}(\delta)$}.

The effect that it is
minimum at $T=T^*_{Re}$ stems from the
behaviour of $\alpha$ which follows the behaviour of
$\chi^0({\bf Q}_{AF}, 0)$, see Fig.9.
To prove the second point let's remind that 
$\chi^0({\bf Q}_{AF}, 0)$, and therefore $\alpha$,
are practically constant for $T<T^*_{Re}$
so that for fixed doping
one has
 $\kappa^2_{exc}(\delta,T) \approx \kappa^2_{exc}(\delta,0)$.
On the other hand, the zero temperature value of $\alpha$ 
(and therefore the $\kappa^2_{exc}(\delta,0)$)
also changes very little when $|\delta-\delta_{exc}(0)|$
changes. It is so in the case when $\delta_{exc}(0)$
is not too close to $\delta_c$ (or $Z_{exc}(0)$
is not too close to $Z=0$, see Fig.10). This condition should be
fulfilled for
 cuprates where
 the interaction
$J$ is extremely strong as discovered experimentally.
It means that $J/W$ should not be small and therefore the
critical value $\delta_{exc}(T=0)$ should be rather far from
$\delta_c$.

 {\bf Conclusion} : {\it The line $T=T^*_{Re}(\delta)$
is the line of almost phase transition. The state below
is characterized
 (i) by a reentrant behaviour as its rigidity increases
with increasing temperature, (ii) by
 small and almost unchanged with doping
$\kappa^2_{exc}$}, i.e. {\bf it is the reentrant
in $T$ and almost frozen in $\delta$ rigid SDW liquid}.
{\it  This
means namely that whatever is doping (even quite far from
$\delta_{exc}(0)$) this state is effectively in a proximity
of the ordered SDW "excitonic" phase.} 

This has very important consequences
when considering the situation
in the presence of SC phase. Since the line
 $T=T^*_{Re}(\delta)$ is always leaning out of the SC phase,
 the normal
state at $T_{sc}(\delta)<T<T^*_{Re}(\delta)$ is quite strange :
{\bf  Being adjacent to the
 SC phase in fact
it corresponds to the proximity of the SDW "excitonic" phase}
\cite{kappasc}. It is this feature which in our opinion is crucial for
understanding the anomalous behaviour in the underdoped
regime of high-$T_c$ cuprates.

On the other hand  $\kappa_{exc}^2$
decreases more or less rapidly with increasing
 the temperature and doping
distances from the line $T=T^*_{Re}(\delta)$
  on
the other side of it 
(see the behaviour of
$\chi^0({\bf Q}_{AF}, 0)$ in Fig.9) so that the state above is a
usual disordered metallic state.

Let's analyze now the behaviour of
the "excitonic" correlation
length in the underdoped regime of the
 metallic state

\begin{equation}
\xi_{exc} = \sqrt A/\kappa_{exc}.
\label{40}
 \end{equation}
As $A$ does not depend on $\delta$ and $T$,
 the behaviour of $\xi_{exc}$ is determined only by the
behaviour of $\kappa^2_{exc}$. 
 Therefore 
$\xi_{exc}$
is maximum at 
$T=T^*_{Re}(\delta)$
(not at $T=0$ as it usually happens in any
quantum disordered state) that is natural in a view that the
line $T=T^*_{Re}(\delta)$ is the line of almost
phase transition.
On the other hand, $A$ is unusually small with respect
 to normal metal where there are two contributions to
$A$ : (i) a dominant one resulting from a $\bf q$
dependence of $Re\chi^0$, (ii) a contribution
coming from the 
$\bf q$
dependence of the interaction. As the former is
absent in our case due to the plateau in the
 $\bf q$
dependence of $Re\chi^0$, $A$ is small. As a result,
$\xi_{exc}$
is not high although $\kappa^2_{exc}$ is small.
This is very untrivial feature: {\bf our rigid SDW
liquid
is not characterized by high correlation length.}
Above the line
$T^*_{Re}(\delta)$
the correlation length decreases
  more or less rapidly in the ordinary way.

To finish the discussion  let's analyze
the behaviour
of the relaxation energy $\omega_0=\kappa^2_{exc}/C$ corresponding
to the imaginary pole of (\ref{34}). Due to the rather strong
increase with $T$ of $C(0)$ in the range $T<T^*_{Im}$ (see Fig.12),
the reentrant behaviour of $\omega_0$ with $T$ is much more
pronounced thaN for  $\kappa^2_{exc}$ and  $\xi_{exc}$ so that
in a proximity of $T^*_{Re}$ the relaxation is very slow.

\subsection{Properties corresponding to those measured by NMR and inelastic neutron
scattering}

All physical properties in the metallic state
are sensitive to the existence
of the line of quasi phase transition
 and we will
progressively consider them in following papers.
 Here we consider only
two examples (for electronic properties see \cite{Kisselev}
and \cite{OPDis}).

Let's firstly consider static and
quasistatic properties
corresponding to measured   $1/T_1T$ and $1/T_{2G}$
on cooper.
The physical characteristics corresponding to them are
$\lim_{\omega \rightarrow 0} Im\chi
({\bf q},\omega)/\omega$ and $Re\chi
({\bf q},0)$ integrated on ${\bf q}$ with some function
peaked in ${\bf Q}_{AF}$. Detailed calculations will be
performed elsewhere. Here we would like to show already
some crude estimations performed in a traditional way.
Since our form (\ref{34}) for the susceptibility in the
limit $\omega \ll \omega_c$ coincides  with the traditional form 

\begin{equation}
\chi_{\bf k}= \frac{\chi_{\bf Q}}{1+k^2\xi^2}
\frac{1}{1-i\omega/\Gamma_{\bf k}}, \hskip 1 cm
\Gamma_{\bf k}=\Gamma_{\bf Q}(1+ k^2\xi^2)
\label{41}
\end{equation}
(where $ k=|{\bf q}-{\bf Q}|$ and ${\bf Q}={\bf Q}_{AF}$) one has
after integration

\begin{equation}
\frac{1}{T_1T} \propto \frac{\chi_{\bf Q}}{\Gamma_{\bf Q}}
\frac{1}{\xi^2}, \hskip 1 cm
\frac{1}{T_{2G}} \propto \frac{\chi_{\bf Q}}{\xi}.
\label{42}
\end{equation}
These are the same expressions which are  used in most
 papers to analyze NMR.
The important difference is that in our case 
the parameters (determined microscopically) are
given by

\begin{equation}
\chi_{\bf Q}=\chi^0_{\bf Q}/\kappa_{exc}^2, \hskip 1 cm
\Gamma_{\bf Q}= \kappa_{exc}^2/C
\label{43}
\end{equation}
and they behave in the anomalous way with changing $T$
due to the anomalous behaviour of $\kappa_{exc}^2$,
$\chi^0_{\bf Q}$, $\xi_{exc}$ and $C$ as discussed above.

Since three parameters, $\chi^0_{\bf Q}$,  $\kappa_{exc}^2$
and  
$\xi$  remain
constant until $T=T^*_{Re}(\delta)$, $\frac{1}{T_{2G}}$
remains constant for $T<T^*_{Re}(\delta)$. As to
$\frac{1}{T_1T}$ it is influenced by two different
temperature dependences :
$\frac{1}{\Gamma_{\bf Q}}$ increases with temperature as
$C(0)$ for $T<T^*_{Im}(\delta)$ and has a maximum at
$T=T^*_{Im}(\delta)$ 
while $\xi$ and $\chi_{\bf Q}$ remain
constant until $T=T^*_{Re}(\delta)$.
As a result we get a picture shown in Fig.19.

\begin{figure}
\begin{center}
\epsfig{%
file=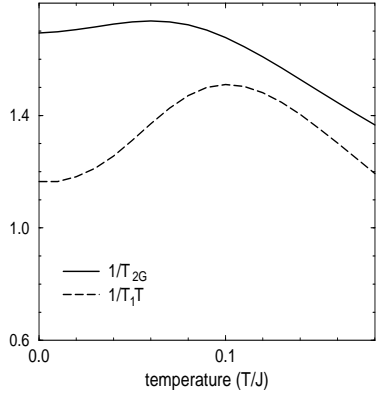,%
figure=figure26.eps,%
height=6cm,%
width=6cm,%
angle=0,%
}
\\
\end{center}
\caption{ Calculated $\frac{1}{T_1T}$ (the dashed line)
 and $\frac{1}{T_{2G}}$ (the solid line)
as functions of $T$ for the underdoped regime
($Z/t=0.19$, $t'/t=-0.3$, $t/J=1.9$
 (the existence of SC state
is ignored).}
\label{f19}
\end{figure}

The plateau in $\frac{1}{T_{2G}}$ exists until 
$T=T^*_{Re}$ while the maximum in
$\frac{1}{T_1T}$ occurs at $T=T^*_{1/T_1T}$.
The latter temperature occurs between $T=T^*_{Re}$
and $T=T^*_{Im}$ since $\frac{1}{T_1T}$ is
sensitive to both $\xi_{exc}$ and $C$. {\bf  In any
case  $T^*_{\frac{1}{T_1T}}>T^*_{\frac{1}{T_{2G}}}=T^*_{Re}$}.

These results explain  well the experimental data
for $\frac{1}{T_{2G}}$ and $\frac{1}{T_1T}$
(quite surprising as underlined by experimentalists),
compare for example Fig.19 with Fig.2 in \cite{Takigawa}. When comparing
one should keep in mind that we ignored the existence of
the SC state when calculating, therefore the theoretical
curves should be considered only above $T_{sc}$.

The existence of the strong AF fluctuations does not mean
that it is only them which always govern the behaviour
of the system. For example there is no reason to think that the
behaviour of $1/T_1T$ on oxygen would be determined 
by the "tail" of these fluctuations. It is determined rather by
fluctuations corresponding to small $\bf q$ which are also
critical in a proximity of the considered
QCP being however not
 enhanced due to the sign of the interaction.
The behaviour of the fluctuations around $\bf q=0$
is completely independent of the behaviour of
AF fluctuations. We will analyse the former fluctuations
 carrefully
elsewhere. Here we only present Fig.20 where we
compare the behaviour of
 $\lim_{\omega \rightarrow 0} Im\chi/\omega$
for $\bf q$ around $\bf q=0$ and around
${\bf Q}_{AF}$ to emphasize their independence :
one can see that  $\lim_{\omega \rightarrow 0} Im\chi/\omega$
grows for $\bf q$ around $\bf q=0$ in the
temperature range where it already
decreases for ${\bf q}={\bf Q}_{AF}$. It is interesting that such a
behaviour
is    very
close to that observed by NMR for
oxygen and cooper \cite{Walstedt}.

\begin{figure}
\begin{center}
\epsfig{%
file=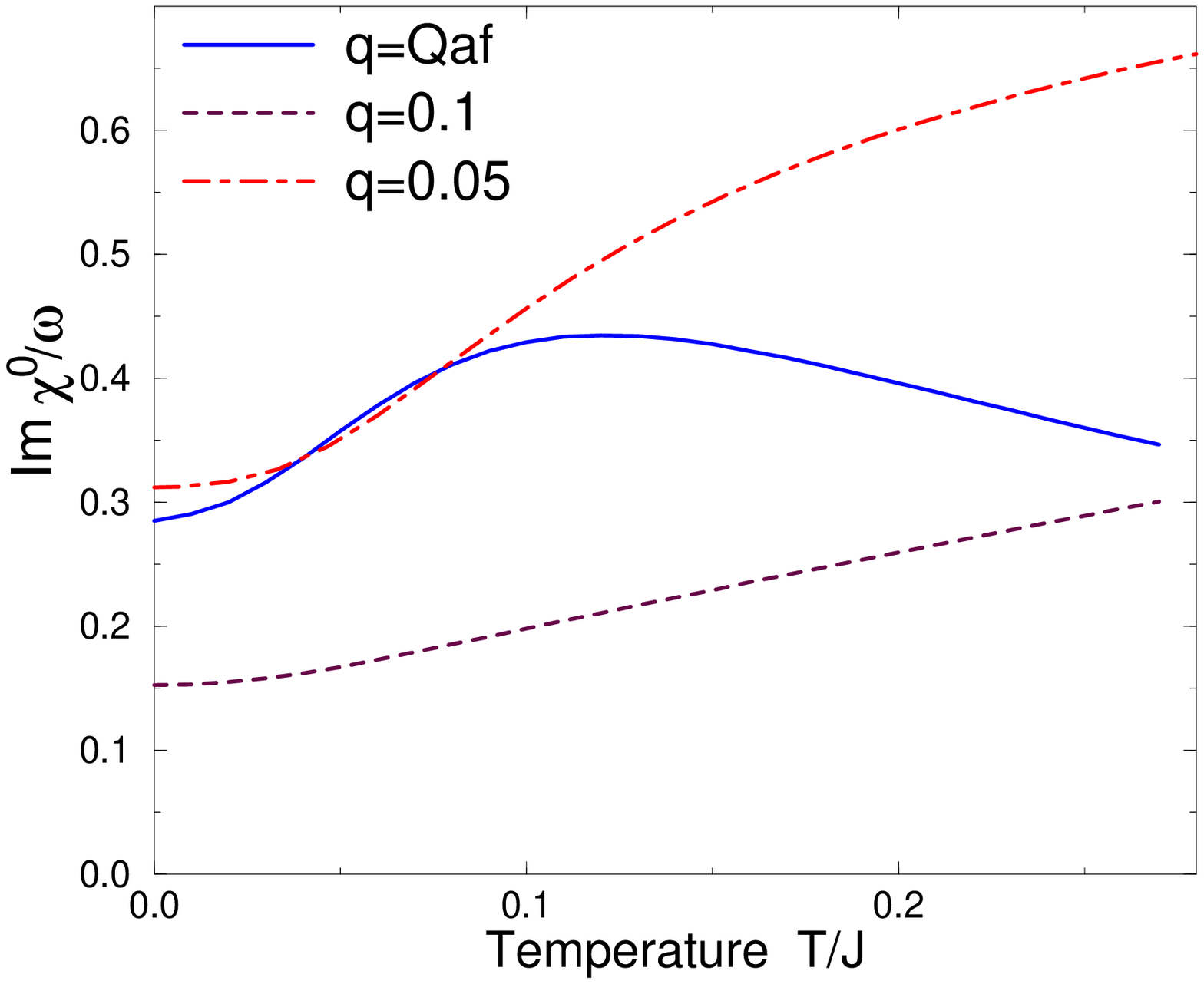,%
figure=figure27.eps,%
height=6cm,%
width=6cm,%
angle=0,%
}
\epsfig{%
file=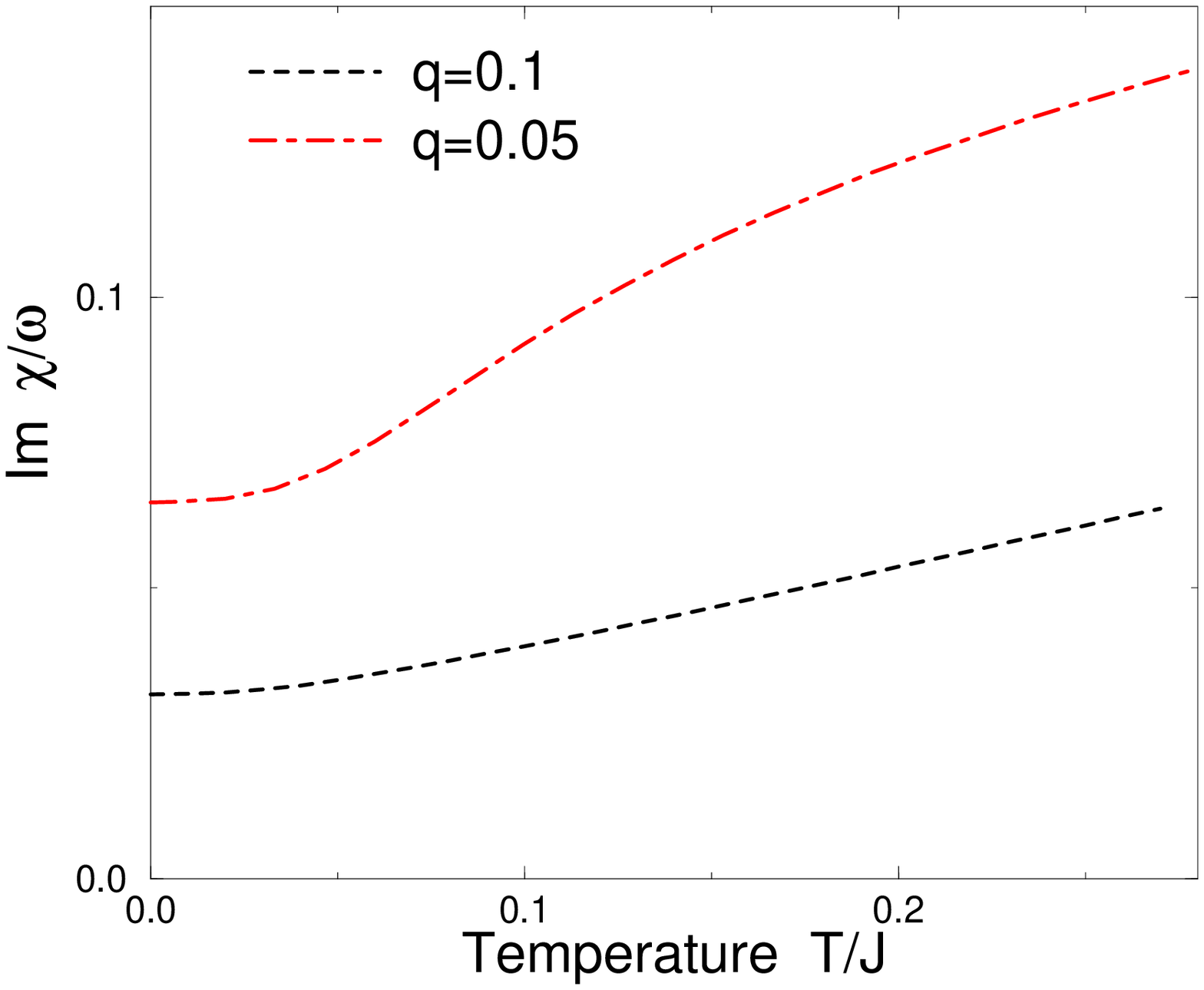,%
figure=figure28.eps,%
height=6cm,%
width=6cm,%
angle=0,%
}
\\
\end{center}
\caption{ Results of numerical
 calculations of  $\lim_{\omega \rightarrow 0}
 Im\chi^0({\bf Q},\omega)/\omega$ as a function
of $T$ for ${\bf q}$ in the vicinity of ${\bf Q}_{AF}$
and in the vicinity of ${\bf q}=0$ (a)
and of  $\lim_{\omega \rightarrow 0}
 Im\chi({\bf Q},\omega)/\omega$ for ${\bf q}$
in the vicinity of ${\bf q}=0$ ($t'/t=-0.3$, $t/J=1.9$,
$Z/t=0.166$)
 (the existence of SC state
is ignored). We can not present the latter function for
both cases at the same graph due to the very different
scale : it is enhanced very stronly in the case
of ${\bf Q}_{AF}$.}
\label{f20}
\end{figure}

Let's  analyze now an $\omega$ dependence of the 
imaginary part of the spin
susceptibility, $Im\chi$, taken around ${\bf q}={\bf Q}_{AF}$
which corresponds to the characteristic measured by neutron
scattering.

  Results of numerical calculations of 
$Im\chi({\bf Q}_{AF},\omega)$ 
as a function of $\omega$ performed based on (\ref{32}) and (\ref{11})
for fixed doping and increasing
 temperature
 are shown in Fig.21. [To analyze the spin dynamics it is
preferable to use the complete form (\ref{32}), (\ref{11}),
 (\ref{1}) rather than the
analytical form (\ref{34}) valid only for $\omega<\omega_c(T)$.]

\begin{figure}
\begin{center}
\epsfig{%
file=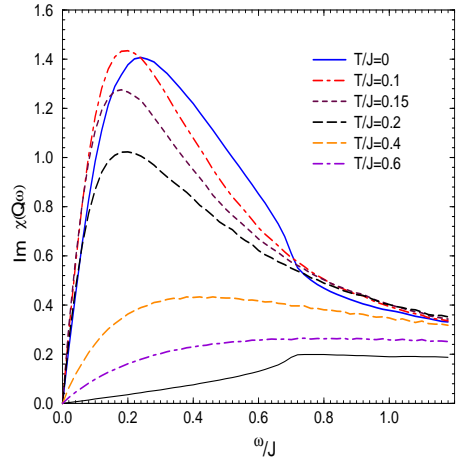,%
figure=figure29.eps,%
height=7cm,%
width=7cm,%
angle=0,%
}
\\
\end{center}
\caption{ Energy dependences of 
 $Im\chi({\bf Q}_{AF},\omega)$ 
for the underdoped regime  for fixed doping 
and increasing temperature.
 For a
comparison we show also the bare susceptibility
 $Im\chi^0({\bf Q}_{AF},\omega)$ for $T=0$ (the thin line).
($t'/t=-0.3$, $t/J=1.83$,
$Z/t=0.29$
For chosen doping $T^*_{Re}/J \approx 0.1$ and
$T^*_{Im}/J \approx 0.2$ ($\delta=0.1$)}
\label{f21}
\end{figure}

Due to the small value of $\kappa^2_{exc}=1-\alpha$
 for $T<T^*_{Re}$ there
is a strong enhancement at low $\omega$ in comparison
with the bare $Im\chi^0$. The position of the low energy peak
is given by

\begin{equation}
\omega=\omega_0 = \frac{\kappa^2_{exc}}{C}
\label{44}
\end{equation}
under the condition $\omega_0 \ll \omega_c$.
This condition is necessary in order $C$ to
not depend on $\omega$. What is important : the
condition $\omega_0 \ll \omega_c$
is fulfilled for all dopings within the
rigid SDW liquid state
 $T<T^*_{Re}(\delta)$
 due to the specific
behaviour of $\kappa^2_{exc}$ discussed in subsection IIIb.

We can distinguish three low temperature regimes
for $Im\chi$.
 The first, for  $T<T_F(\delta)$, is the regime where both
$\kappa^2_{exc}$ and $C(\omega_0) \approx C(0)$
do not depend on $T$. $T_F$ is very low 
(see Fig.12)
so that this
regime   can hardly be detectable. The second regime, for
 $T_F(\delta)<T<
T^*_{Re}(\delta)$, {\bf corresponding  to the 
reentrant rigid SDW liquid state} is the regime where
 $\kappa^2_{exc}$ slightly decreases with $T$
 while  $C(\omega_0) \approx C(0)$
 increases with  $T$.
 As a result  in this regime  the energy
of the peak, $\omega_0$,
{\bf decreases with increasing $T$} (reentrant behaviour). The third
 regime, occuring 
for $T^*_{Re}<T<T^*_{Im}$ (i.e. {\bf within the zone of quasi
phase transition}) is the regime 
in which  $\kappa^2_{exc}$ starts to increase with $T$
 while $C(\omega_0)$ continues to grow. As a result in this regime
 the energy of the peak, $\omega_0$,
{\bf is unchanging with increasing $T$}.

Both latter behaviours are anomalous. For an usual
metal considered in a critical regime
above a magnetic phase transition,
$\kappa^2_{exc}$ decreases with 
increasing $T$ and $C$ is constant as
$T$ increases  that leads to the ordinary behaviour
with the {\bf low energy peak moving towards high energies with
increasing $T$}. For the system under consideration
 such a behaviour occurs in the  regime,
$T>T^*_{Im}$, i.e. {\bf after
passing the zone of quasi phase transition}.

The value of $Im\chi$ at the peak position is
equal to $4J Im\chi({\bf Q}_{AF},\omega_0)=1/2\kappa^2_{exc}$.
It practically does not change 
with $T$ in the first regime, even slightly increases
with increasing $T$. For the second and third regimes,
 $T>T^*_{Re}$,
 $Im\chi({\bf Q}_{AF},\omega_0)$ decreases with
increasing $T$ in the ordinary way.

Is it possible to observe experimentally the discussed above
 anomalous regimes ?
It is difficult because the low temperature regimes
are hidden under the SC phase. Nevetheless it is
possible for very underdoped materials. [{\bf Another
possibility would be a neutron experiment in a pulsed
magnetic fields]}. Although INS data 
in the normal state are in general
in a good agreement with our
results (see for example \cite{Regnault}
there is no data  available concerning the behaviour of
 $Im\chi({\bf Q}_{AF},\omega)$ as a function of $\omega$
  obtained in progressive changing $T$ for temperature range
just above $T_c$. 
For example the recent data
 for the normal state  in
highly underdoped 
YBCO \cite{Keimer} correspond to $60K$ and to $200K$, i.e.
there is an important jump in temperature,
probably across different regimes.
{\bf To verify the
predicted  anomalous behaviour it is desirable
to perform such measurements in 
progressive  changing $T$}. In fact our proposal is a way
to observe $T^*(\delta)$ by INS (as it is still one of a few
 experiments where $T^*(\delta)$ has not yet
been observed).


\section{Summary and discussion}

The results obtained in the paper  have two aspects :
one fundamental and one concerning  the high-$T_c$
cuprates.

{\bf The fundemental aspect} is following. 
We have shown
that a 2D system of noninteracting electrons on 
a square lattice with hoping between more than nearest neighbours
undergoes   a specific
 electronic topological
transition (ETT) at some
electron concentration $\delta=\delta_c$. 
The point of ETT is a $T=0$ quantum critical
point with  several characteristic aspects.
The first aspect is related to the local change of topology
of the FS near SP's. This results in
 singularities in thermodynamic
properties, in  a ferromagnetic  response function, in an
 additional divergence of the
  superconducting
 response function. From this point of view
it is a QCP of a gaussian type.
 [The  logarithmic singularity in a
density of states at $\omega=0$ is a consequence of this
 aspect of ETT not a reason].
The other aspects are related  to the topological change 
at $\delta=\delta_c$ in 
 mutual properties of the FS in the vicinities of two 
different SP's. As a result, the behaviour of the system
is very asymmetrical on two sides of the ETT, the point
$\delta=\delta_c$ appears {\it as
 the end
of the critical line of   static
Kohn singularities} in the polarizability of noninteracting
electrons which exists on one side of QCP, $\delta>\delta_c$.
On the other side, $\delta<\delta_c$, Kohn singularity manifests
 itself as the
line of the dynamic singularities. The  dynamic singularities
at $T=0$ are transforming into static  anomalies at finite
temperature. All this happens for the characteristic
wavevector for this regime  ${\bf q=Q}_{AF}$.

The specific behaviour of the system of noninteracting electrons
related to ETT leads to the anomalous behaviour
of the system in the presence of interaction. The anomalies
exist whatever is a type of interaction : in a triplet or
singlet channels,  $\bf q$ dependent or $\bf q$ independent,
since the motor for them is the ETT in the noninteracting system.
We study some of the anomalies. For example we show that the
line of instability of the initial metallic state against SDW
or CDW order (depending on the type of interaction) has the
anomalous form : it grows from the QCP with increasing
the distance from QCP instead of having the ordinary form of
the bell around QCP.
We show that in the metallic state out of this phase 
there is a characteristic temperature
for each electron concentration, $T^*_{Re}(\delta)$, which is 
 the temperature
of an almost phase transition : a rigidity of the metallic
state and the correlation length are maximal
at $T=T^*_{Re}(\delta)$ when approaching from above
and from below.
We show that the state below $T=T^*_{Re}(\delta)$ (and out of the
ordered phase) is characterized by a very specific behaviour.
It is almost frozen in distance from
QCP and reentrant in temperature  rigid SDW (or CDW) liquid. 
We study some properties of this state.
[Some details of the anomalous behaviour depend on interaction. When
analysing we have
choosen to consider the $\bf q$ dependent interaction
in a triplet channel in a view of the following
applications to high-$T_c$ cuprates.]

All these anomalies happen on one side of QCP, 
$\delta<\delta_c$. On the other side,
 $\delta>\delta_c$, the behaviour is quite ordinary.
For example, the correlation length decreases monotonically
and quite rapidly with $T$ and with the distance from QCP,
$|\delta-\delta_c|$, the  line of of SDW or CDW "excitonic"
instability
behaves in the ordinary way etc.

To finish with the fundamental aspect we would like
to emphasize that the
considered  ETT is quite general and exists in all cases of hoping between
 more than nearest neighbours: $t' \neq 0$ or/and $t'' \neq 0$
etc. For any set of these parameters one can introduce
an effective $\tilde{t}'$ and map the situation into the
considered in the paper  generic model. The exceptions
are some sets \cite{OPQCP2}, including $t'=t''=...=0$, for which
the last aspect of the ETT disappears : the QCP is no
more the end of the line of static Kohn singularities, 
 the behaviour is absolutely symmetrical
on two sides of  $\delta=\delta_c$.  This case corresponds
 to the nested FS.
 And although the
first aspect of criticality still exists (with the
"famous" Van Hove singularity scenario) all discussed
in the paper anomalies disappear. We emphasize this again
 to avoid a misunderstanding : the considered in
the paper scenario has nothing to do with the
 Van Hove singularity.

{\bf The second aspect of the paper} is an application
of the theory  to
the hole-doped high-$T_c$ cuprates. These materials are quasi-2D
systems, the electron FS observed experimentally has
such a shape which implies the existence of hoping $t'$
(or/and $t''$, etc.), the shape of FS changes continuously
towards the form corresponding to the ETT when moving
from the underdoped side towards the optimal for
superconductivity doping. On the other hand, the
observed experimentally (by INS and NMR) strong
spin dependent
 response around ${\bf q}={\bf Q}_{AF}$
is a phenomenological argument in a favour of the strong
momentum dependent interaction in a triplet
channel. 

All these features allow us to apply the theory
to the  high-$T_c$ cuprates and to
consider the discussed scenario of the anomalous
behaviour
as a very probable origin of
the anomalies observed in the underdoped regime
of the hole-doped high-$T_c$ cuprates.

In the present paper we have considered some
properties and have compared them with experiments.
The most important result is of course the existence of the
generic temperatures $T^*_{Re}(\delta)$ and $T^*_{Im}(\delta)$
which give rise to the existence in the metallic
state out of the ordered phase
of the
characteristic temperatures $T^*$, different for
different properties, but always proportional to 
$\delta_c-\delta$. This is in a good agreement
with the general situation in the cuprates. We have
considered some concrete examples. We analyse
the behaviour of the physical characteristics corresponding to
the nuclear spin
 lattice relaxation rate $1/T_1T$  and to the nuclear transverse
relaxation rate
$1/T_{2G}$  on copper. We have shown that the
latter  is almost constant as a function of $T$ until
some temperature 
$T=T^*_{1/T_{2G}}$ and then decreases with $T$.
On the other hand, the first function increases with
increasing $T$ until $T^*_{1/T_1T}$ and then
decreases. The characteristic temperatures are
different and $T^*_{1/T_1T}>T^*_{1/T_{2G}}$.
All these features explain quite well the experimentally
observed behaviour, see for example \cite{Takigawa}.
We show that the behaviour of the spin responce function is
quite different in the cases of $\bf q$ around ${\bf Q}_{AF}$
and  $\bf q$ around $\bf q=0$ being completely independent.  
Fluctuations corresponding to small $\bf q$  are also
critical in a proximity of the ETT being however not
 enhanced due to the sign of the interaction. This can explain
the observed experimentally qualitatively
different behaviout of $1/T_1T$ on
oxygen and cooper \cite{Walstedt}.
We have analyzed briefly  the behaviour of
the characteristics corresponding to
that measured by INS and we have demonstrated how $T^*$ can be seen in
neutron scattering experiment.
As to the correlation length there are at present two
 contradictory
conclusions \cite{Bobroff,Slichter} about its behaviour
 based on the same
measurements (performed unfortunetely starting from
quite high temperature). Since the method
 is indirect, an answer depends crucially on the
model for the $\bf q$-dependence used to extract $\xi$.
In both cases \cite{Bobroff,Slichter} the used
models are standard : the lorentzian and the gaussian.
As we have seen, the  $\bf q$ dependence can be quite
nonstandard so that one should be very cautious when
interpretating the experiment. As to INS data, they give
 only a $\bf q$ width  of $Im\chi$ at finite $\omega$
not a $\bf q$ width  of $Re\chi$ at $\omega=0$ which
corresponds to the correlation length. And
although the INS data show
a  $\bf q$ width   not depending on $T$
(that indirectly  can be considered as an argument in a favour of our theory)
it can not be considered as a crucial experiment for $\xi$.

More detailed analysis of NMR and INS will be performed
elsewhere. Electronic properties in the ordered SDW "excitonic"
phase
and in the reentrant SDW liquid state are considered and compared
with ARPES in \cite{Kisselev,OPDis} respectively.

\end{fmffile}


\begin{thebibliography} {33}

\bibitem{NMR} H.Alloul,T.Ohno, P.Mendels,
Bull.Am.Phys.Soc.  {\bf 34}, 633 (1989);
Phys.Rev.Lett. {\bf 63}, 1700 (1989);
W.W.Warren et al, Phys.Rev.Lett. {\bf 62}, 1193 (1989);
G.V.M.Williams et al Phys.Rev.Lett. {\bf 78}, 721 (1997);

\bibitem{Takigawa} M. Takigawa,  Phys.Rev.B
{\bf 49}, 4158 (1994)



\bibitem{ARPES} H.Ding, T. Yokoya, J.C. Campuzano et al,
Nature (London), {\bf 382}, 51 (1996);
 H.Ding, J.C. Campuzano, M.R. Norman, cond-mat/9712100

\bibitem{conductivity} S.L. Cooper et al, Phys.Rev.B
{\bf 40}, 11358 (1989); Puchkov et al,
Phys.Rev.Lett. {\bf 77}, 3212 (1996)  

\bibitem{transport}H.Y.Hwang et al, Phys.Rev.Lett. {\bf 72}, 2636 (1994)

\bibitem{thermopower}J.L.Talon, J.R. Cooper, P.S.I.P.N. de Silva
et al, Phys.Rev.Lett. {\bf 75}, 4114 (1995)

\bibitem{heatcapacity} J.W. Loram et al, 
 Phys.Rev.Lett. {\bf 71}, 1740 (1993)



\bibitem{susc}D.C. Johnston,  Phys.Rev.Lett. 
{\bf 62}, 957 (1989)

\bibitem{Raman} R. Nemetschek et al, Phys.Rev.Lett.
 {\bf 78}, 4837 (1997)

\bibitem{Kisselev} F.Onufrieva, M. Kisselev, P.Pfeuty, to
be published


\bibitem{OPDis} F.Onufrieva, P.Pfeuty, to
be published


\bibitem{Van-Hove}L. Van Hove,  Phys.Rev.
{\bf 9}, 1189 (1953). The existence of the logarithmic
 singularity
in the density of states for 2D case was
 first shown by E. Montroll \cite{Montroll}
for the case of phonon density of states.

\bibitem{Montroll} E. Montroll, J.Chem.Phys.
{\bf 15}, 575 (1947)
 

\bibitem{Kohn}W.Kohn Phys.Rev.Lett. {\bf 2}, 393 (1959)

\bibitem{Kohn2}L.Roth, H.J. Zeiger, T.A. Kaplan,
 Phys.Rev.
{\bf 149}, 519 (1966) 

\bibitem{Rice}T.M. Rice,  Phys.Rev.B
{\bf 2}, 3619 (1970)


\bibitem{neighbours}More presisely, it exists
in all cases $t' \neq 0$ or/and $t'' \neq 0$ etc. except
for the special set of the parameters (including
the case $t'=t''=...\rightarrow 0$) corresponding to the perfect nesting
of FS. 



\bibitem{Lifshitz}
 I.M. Lifshitz, 
Zh. Eskp. Teor. Fiz. {\bf 33}, 1569 (1960)

\bibitem{ETT} A.A. Varlamov, 
V.S. Egorov, A. Pantsulaya, Adv. in Phys. {\bf 38
}, 465 (1989)

\bibitem{OPQCP2} F.Onufrieva, P.Pfeuty, to
be published

\bibitem{Walstedt} R.E. Walstedt, B.S. Shastry, S.W.Cheong,
 Phys.Rev.Lett.
 {\bf 72}, 3610 (1994)

\bibitem{OPINS} F.Onufrieva, P.Pfeuty, to
be published





\bibitem{excitonic} B.I. Halperin, T.M. Rice, Solid State Phys.,
{\bf 21}, 115 (1968); A.Kozlov, L.Maximov, Sov.Phys.JETP,
{\bf 21}, 790 (1965)


\bibitem{scaling}Scaling is the main property of a quantum critical
point. In Sec.II.C we gave scaling forms for 
 $Im\chi^0({\bf Q}_{AF},\omega,Z,T=0)$ and 
 $Re\chi^0({\bf Q}_{AF},\omega,Z,T=0)$. The functions
 $Re\chi^0({\bf Q}_{AF},\omega=0,Z,T)$ and $\lim_{\omega \rightarrow 0}
Im\chi^0({\bf Q}_{AF},\omega,Z,T)/\omega$ can also be shown to scale
with the scaling variable $T/Z$ and a very anomalous scaling
function. This point will be discussed in another paper.



\bibitem{OnufrievaPRB96} F.Onufrieva, S.Petit, Y.Sidis
Phys.Rev.B
{\bf 54}, 12464 (1996)

\bibitem{finiteT}No true finite $T$ ordered state is possible in 2D system
in the case of continuous symmetry of the Hamiltonian and
only a possible   Kosterlitz-Thouless transition
is
expected. RPA finite $T$ transitions should be
interpreted for pure 2D system as crossover signaling
the appearence of large but finite correlation length.
However a weak 3D coupling or small anisotropy of
the Hamiltonian are sufficient to induce true LRO.

\bibitem{susceptibility}The form (\ref{34}) is close to the phenomenological
susceptibility introduced by A.Millis et al \cite{Pines}
if one considers $C=const$. [This form has been used in
many following papers by D.Pines et al (see for example
\cite{pines}) and A.Chubukov et al (see for example
\cite{Chubukov})]
As we have shown this form is valid only for
very low $\omega$, $\omega \ll \omega_c \propto Z$.
For higher  $\omega$, there is an effect of the
effective "pseudogap". We mean  that
 $C$ is no more constant when  $\omega$ is not
extremely small, see Fig.8. Another important difference is
that this form represents only fluctuations in
 the vicinity of ${\bf Q}_{AF}$. As we have shown in Sec.II there is
also a singularity of $Re\chi^0({\bf q},0)$ in
 the vicinity of ${\bf q}=0$ and the
 fluctuations related to that. Their behaviour is
very different from the behaviour of fluctuations
 in
 the vicinity of ${\bf Q}_{AF}$ that leads to
an independent behaviour of $\chi^0({\bf q}=0,0)$
and $\chi^0({\bf q}={\bf Q}_{AF},0)$ as functions of $T$
and $Z$. And finally and most important is that the
parameters in the form (\ref{34}) behave in a very untrivial way
 as functions of $T$ and $\delta$ as discussed in the text.



\bibitem{Pines}A. Millis, H. Monien, D. Pines,  Phys.Rev.B
{\bf 42}, 167 (1990)
 
\bibitem{pines}V. Barzykin, D.Pines,  Phys.Rev.B
{\bf 52}, 13585 (1995)

\bibitem{Chubukov}A. Chubukov, D. Morr, cond-mat/9701196


\bibitem{kappasc}Analysis of the behaviour
of the parameter $\kappa^2_{sc}=1-V^{sc}\Pi^{sc}_d(0,0)$
describing a proximity to SC phase
is  performed in details in \cite{OPINS}; however
it is clear from the breaf analysis performed in the Subsec.IIE
that it behaves in the ordinary way, i.e. increases
quite rapidly when one goes away from the critical
line $T_{sc}(\delta)$.
Therefore, except for an intime vicinity of $T_{sc}(\delta)$,
it is valid : $\kappa^2_{sc}>\kappa^2_{exc}$.

\bibitem{Regnault}L.P. Regnault, P. Bourges, P. Burlet et al,
Physica C {\bf 235-240}, 59 (1994)


\bibitem{Keimer} P. Bourges, H.F. Fong, L.P. Regnault
et al,  Phys.Rev.B
{\bf 56}, R11439 (1997)







\bibitem{Bobroff}J. Bobroff, H. Alloul, Y. Yoshinari,
 Phys.Rev.Lett.
 {\bf 79}, 2117 (1997)




\bibitem{Slichter}D. Morr, J. Schmalian, R. Stern, C.P. Slichter,
cond-mat/9801317


\end{thebibliography}
\end{document}